\documentclass[
journal
]{IEEEtran}

\usepackage{amsmath}
\usepackage{amssymb}
\usepackage{amsfonts}
\usepackage{amsthm}
\usepackage{array}
\usepackage{cite}
\usepackage{algorithm}
\usepackage{algpseudocode}
\usepackage{float}
\usepackage{tikz}

\newtheorem{theorem}{Theorem}

\newtheorem{lemma}{Lemma}
\newtheorem{corollary}{Corollary}
\newtheorem{definition}{Definition}

\newcommand{\bcomment}[1]{}
\newcommand{\TODO}[1]{}
\newcommand{\R}{\mathbb{R}}
\newcommand{\N}{\mathbb{N}}
\newcommand{\one}{\mathbf{1}}
\newcommand{\zero}{\mathbf{0}}
\newcommand{\two}{\{0,1\}}
\newcommand{\mtrx}[3]{{#1}^{#2 \times #3}}
\newcommand{\deq}{=}%{\triangleq}
\newcommand{\diag}[1]{\operatorname{diag}(#1)}
\newcommand{\lp}[4]{%
\begin{IEEEeqnarray*}{rCrl}
#1 &=& #2                & #3\\ 
   & &\text{\rm s. t.}\; & #4
\end{IEEEeqnarray*}
}
\newcommand{\lpp}[6]{%
\begin{IEEEeqnarray*}{rCrl}
#1 &#2& #3                & #4\\ 
   &  &\text{\rm s. t.}\; & #5\\
   &  &                   & #6
\end{IEEEeqnarray*}
}
\DeclareMathOperator*{\E}{\mathbb{E}}
\DeclareMathOperator*{\argmin}{argmin}

\begin{document}
\title{Generalized sphere-packing and sphere-covering bounds on the size of codes for combinatorial channels}
\author{Daniel~Cullina,~\IEEEmembership{Student Member,~IEEE}
        and Negar~Kiyavash,~\IEEEmembership{Senior Member,~IEEE}%
\thanks{The material in this paper was presented (in part) at the International Symposium on Information Theory, Honolulu, July 2014~\cite{cullina_improvement_2013}.
This work was supported in part by NSF grants CCF 10-54937 CAR and CCF 10-65022 Kiyavash.}%
\thanks{Daniel Cullina is with the Department of Electrical and Computer Engineering and the Coordinated Science Laboratory, University of Illinois at Champaign-Urbana, Urbana, Illinois 61801 (email: cullina@illinois.edu). }%
\thanks{Negar Kiyavash is with the Department of Industrial and Enterprise Systems Engineering and the Coordinated Science Laboratory, University of Illinois at Champaign-Urbana, Urbana, Illinois 61801 (email: kiyavash@illinois.edu). }%
}

\maketitle
\begin{abstract}
Many of the classic problems of coding theory are highly symmetric, which makes it easy to derive sphere-packing upper bounds and sphere-covering lower bounds on the size of codes.
We discuss the generalizations of sphere-packing and sphere-covering bounds to arbitrary error models.
These generalizations become especially important when the sizes of the error spheres are nonuniform.
The best possible sphere-packing and sphere-covering bounds are solutions to linear programs.
We derive a series of bounds from approximations to packing and covering problems and study the relationships and trade-offs between them.
We compare sphere-covering lower bounds with other graph theoretic lower bounds such as Tur\'{a}n's theorem.
We show how to obtain upper bounds by optimizing across a family of channels that admit the same codes.
We present a generalization of the local degree bound of Kulkarni and Kiyavash and use it to improve the best known upper bounds on the sizes of single deletion correcting codes and single grain error correcting codes.
\end{abstract}

\section{Introduction}
The classic problem of coding theory, correcting substitution errors in a vector of $q$-ary symbols, is highly symmetric.
First, if $s$ errors are required to change a vector $x$ into another vector $y$, then $s$ errors are also required to change $y$ into $x$.
Second, the number of vectors that can be produced from $x$ by making up to $s$ substitutions, the size of the sphere around $x$, does not depend on $x$.
The sizes of these spheres play a crucial rule in both upper and lower bounds on the size of the largest $s$-substitution-error-correcting codes.
The Hamming bound is a sphere-packing upper bound and the Gilbert-Varshamov lower bound is a sphere-covering lower bound.
The two symmetries that we have described make the proofs of the Hamming and Gilbert-Varshamov bounds extremely simple.

Many other interesting error models do not have this degree of symmetry.
Substitution errors with a restricted set of allowed substitutions are sometimes of interest.
The simplest example is the binary asymmetric errors, which can replace a one with a zero but cannot replace a zero with a one.
Binary asymmetric errors have neither of the two symmetries we have described.
Erasure and deletion errors differ from substitution errors in a more fundamental way: the error operation takes an input from one set and produces an output from another.

In this paper, we will discuss the generalizations of sphere-packing and sphere-covering bounds to arbitrary error models.
These generalizations become especially important when the sizes of the error spheres are nonuniform.
Sphere-packing and sphere-covering bounds are fundamentally related to linear programming and the best possible versions of the bounds are solutions to linear programs.
In highly symmetric cases, including many classical error models, it is often possible to get the best possible sphere-packing bound without directly considering any linear programs.
For less symmetric channels, the linear programming perspective becomes essential.

In fact, recently a new bound, explicitly derived via linear programming, was applied by Kulkarni and Kiyavash to find an upper bound on the size of deletion-correcting codes \cite{kulkarni_nonasymptotic_2013}.
It was subsequently applied to grain errors \cite{kashyap_upper_2013, gabrys_correcting_2013} and multipermutation errors \cite{buzaglo_error-correcting_2013}.
We will refer to this as the local degree bound.
The local degree bound constructs a dual feasible point for the sphere-packing linear program because computation of the exact solution is intractable. Deletion errors, like most interesting error models, act on an exponentially large input space.
Because computation of the best possible packing and covering bounds is often intractable, simplified bounds such as the local degree bound are useful.

Sphere-packing and sphere-covering arguments have been applied in an ad hoc fashion throughout the coding theory literature.
We attempt to present a unifying framework that that permits such arguments in their most general form applicable to both uniform and nonuniform error sphere sizes. 
More precisely, we derive a series of bounds from approximations to packing and covering problems. 
The local degree bound of  \cite{kulkarni_nonasymptotic_2013} is one of the bounds in the series. 
We associate each bound with an iterative procedure such that the original bound is the result of a single step.
This characterization makes it easy to study the relationships between the bounds.
We apply our generalization of the local degree bound to improve the best known upper bounds on the sizes of single deletion correcting codes and single grain error correcting codes.

We use the concept of a combinatorial channel to represent an error model in a fashion that makes the connection to linear programming natural.
These bounds use varying levels of information about structure of the error model and consequently make trade-offs between performance and complexity. 
For example, one bound uses the distribution of the sizes of spheres in the space while another uses only the size of the smallest sphere.

In general, there are many different combinatorial channels that admit the same codes.
However, each channel gives a different sphere-packing upper bound.
We show that the Hamming bound, which can be derived from a substitution error channel, the Singleton bound, which can be derived from an erasure channel, and a family of intermediate bounds provide an example of this phenomenon.

Sphere-packing upper bounds and sphere-covering lower bounds are not completely symmetric.
The linear programming approximation techniques that we discuss each yields a sphere-packing upper bound and a sphere-covering lower bound. 
We explore the relationship between sphere-covering lower bounds and other graph theoretic lower bounds such as Tur\'{a}n's theorem.

Our contributions can be summarized as follows.
We provide a unified framework for describing both upper and lower bounds on code size.
This allows us to make very general statements about the relative strengths of the bounds.
In particular, our generalization of the local degree bound allows us to improve the best known upper bounds for a few channels.
Finally, we demonstrate the power of considering families of channels with the same codes.

In Section~\ref{section:lp-bounds}, we discuss the linear programs associated with sphere-packing bounds.
In Section~\ref{section:iterative}, we present a generalization of the local degree bound that is related to an iterative procedure.
We use this to improve the best known upper bounds on the sizes of single deletion correcting code and single grain error correcting codes.
In Section~\ref{section:deg-seq}, we discuss sphere-packing bounds related to the degree sequence and average degree of a channel.
In Section~\ref{section:families}, we discuss families of channels that have the same codes but give different sphere-packing bounds.
In Section~\ref{section:lower-bounds}, we discuss sphere-covering lower bounds, lower bounds related to Tur\'{a}n's theorem, and their relationships.

\section{Sphere-Packing Bounds and Linear Programs}
\label{section:lp-bounds}
\subsection{Notation}
Let $X$ and $Y$ be a finite sets.
For a semiring $R$, let $R^X$ denote the set of $|X|$-dimensional column vectors of elements of $R$ indexed by $X$.
Let $\mtrx{R}{X}{Y}$ denote the set of $|X|$ by $|Y|$ matrices of elements of $R$ with the rows indexed by $X$ and the columns indexed by $Y$.
Let $2^X$ denote the power set of $X$.
Let $\N$ denote the set of nonnegative integers and let $[n]$ denote the set of nonnegative integers less than $n$: $\{0,1,\ldots,n-1\}$.
Let $\one$ be the column vector of all ones and let $\zero$ be the column vector of all zeros.
For a set $S \subseteq X$, let $\one_S \in \{0,1\}^X$ be the indicator column vector for the set $S$.

\subsection{Combinatorial channels}
We use the concept of a combinatorial channel to formalize a set of possible errors.
\begin{definition}
A combinatorial channel is a matrix $A \in \mtrx{\two}{X}{Y}$, where $X$ is the set of channel inputs and $Y$ is the set of channel outputs.
An output $y$ can be produced from an input $x$ by the channel if $A_{x,y} = 1$.
Each row or column of $A$ must contain at least a single one, so each input can produce some output and each output can be produced from some input.
\end{definition}

We will often think of a channel as a bipartite graph.
In this case, the left vertex set is $X$, the right vertex set is $Y$, and $A$ is the bipartite adjacency matrix.
We will refer to this bipartite graph as the \emph{channel graph}.\footnote{
An equivalent approach, taken by Kulkarni and Kiyavash \cite{kulkarni_nonasymptotic_2013}, is to represent an error model by a hypergraph.
A hypergraph consists of a vertex set and a family of hyperedges.
Each hyperedge is a nonempty subset of the vertices.
A hypergraph $\mathcal{H} = (V,E)$ can be described by a vertex-hyperedge incidence matrix $H \in \mtrx{\two}{V}{E}$.
There are two ways to encode as error model as a hypergraph.
Let $A \in \mtrx{\two}{X}{Y}$ be the combinatorial channel for that error model.
The first option is to take $H = A$ to be the incidence matrix of the hypergraph.
The hypergraph vertices and the channel inputs and there is an edge for each output.
Alternatively, we can let $H = A^T$ and obtain the dual of the previous hypergraph.
Now the hypergraph vertices are the channel outputs.
This is the option taken by Kulkarni and Kiyavash.

Throughout this paper, we use the language of channels and bipartite channel graphs rather than that of hypergraphs.
This allows us to refer to channel inputs and outputs using symmetric language and avoids any confusion between a hypergraph and its dual.
}
For $x \in X$, let $N_A(x) \subseteq Y$ be the neighborhood of $x$ in the channel graph (the set of outputs that can be produced from $x$).
The degree of $x$ is $|N_A(x)|$.
For $y \in Y$, let $N_A(y) \subseteq X$ be the neighborhood of $y$ in the channel graph (the set of inputs that can produce $y$).
In most cases, the channel involved will be evident and we will drop the subscript on $N$. 

Note that $A\one_{\{y\}} = \one_{N(y)}$ and $\one_{\{x\}}^T A = \one_{N(x)}^T$.
Thus $A\one$ is the vector of input degrees of the channel graph, $A^T\one$ is the vector of output degrees, and $\one^TA\one$ is the number of edges.

We are interested in the problem of transmitting information through a combinatorial channel with no possibility of error.
To do this, the transmitter only uses a subset of the possible channel inputs in such a way that the receiver can always determine which input was transmitted.
\begin{definition}
A code for a combinatorial channel $A \in \mtrx{\two}{X}{Y}$ is a set $C \subseteq X$ such that for all $y \in Y$, $|N(y) \cap C| \leq 1$.
\end{definition}
This condition ensures that decoding is always possible: if $y$ is received, the transmitted symbol must have been the unique element of $N(y) \cap C$.

\subsection{Sphere-packing}
\label{section:packing}
A code is a packing of the neighborhoods of the inputs into the output space.
The neighborhoods of the codewords must be disjoint and each neighborhood contains at least $\min_{x \in X} |N(x)|$ outputs.
Thus the simplest sphere-packing upper bound on the size of a code $C$ is
\begin{equation*}
|C| \leq \frac{|Y|}{\min_{x \in X} |N_A(x)|}.
\end{equation*}
This is the \emph{minimum degree upper bound}, because $|N_A(x)|$ is the degree of $x$ in the channel graph of $A$.
The sphere-packing upper bounds discussed in this paper are generalizations of and improvements on this bound.

Maximum input packing and its dual, minimum output covering, are naturally expressed as integer linear programs.
\begin{definition}
\label{def:lp}
For a channel $A \in \mtrx{\two}{X}{Y}$, the size of the largest input packing, or code, is
\lp{p(A)}{\max_{w \in \N^X}}{\one^T w}{A^T w \leq \one.}
The size of the smallest output covering is 
\lp{\kappa(A)}{\min_{z \in \N^Y}}{\one^T z}{A z \geq \one.}
\end{definition}
An output covering can be thought of as a strategy for the adversary operating the channel that is independent of the transmitter's choice of code.
In Section~\ref{section:lower-bounds}, we use output coverings of an auxiliary channel to obtain a lower bound on the $p(A)$.

\subsection{Fractional relaxations}
However, 
%the maximum vertex packing and the minimum hyperedge covering problems over the class of general hypergraphs are NP-Hard.
the maximum independent set and minimum dominating set problems over general graphs are NP-Hard \cite{west_introduction_2001}.
The approximate versions of these problems are also hard.
The maximum independent set of an $n$-vertex graph cannot be approximated within a factor of $n^{1-\epsilon}$ for any epsilon unless P=NP \cite{hastad_clique_1996}.
We seek efficiently computable bounds.
These bounds cannot be good for all graphs, but they will perform reasonably well for many of the graphs that we are interested in.

The relaxed problem, maximum fractional set packing, provides an upper bound on the original packing problem.
\begin{definition}
Let $A \in \mtrx{\two}{X}{Y}$ be a channel.
The size of the maximum fractional input packing in $A$ is
\lpp{p^*(A)}{=}{\max_{w \in \R^X}}{\one^T w}{w \geq \zero}{A^T w \leq \one .}
The size of the minimum fractional output covering is
\lpp{\kappa^*(A)}{=}{\min_{z \in \R^Y}}{\one^T z}{z \geq \zero}{A z \geq \one .}
\end{definition}
The fractional programs have larger feasible spaces, so $p(A) \leq p^*(A)$ and $\kappa^*(A) \leq \kappa(A)$.
By strong linear programming duality, $p^*(A) = \kappa^*(A)$.

Unlike the integer programs, the values of the fractional linear programs can be computed in polynomial time.
However, we are usually in sequences of channels with exponentially large input and output spaces.
In these cases, finding exact solutions to the linear programs is intractable but we would still like to know as much as possible about the behavior of the solutions.

\section{The local degree iterative algorithm}
\label{section:iterative}
Let $A \in \mtrx{\two}{X}{Y}$ be a channel.
We can obtain an upper bound for $p^*(A)$ (and consequently $p(A)$) by finding a feasible point in the program for $\kappa^*(A)$.
Given some $t \in \R^Y$, consider the smallest $c \in \R$ such that $z = ct$ is feasible for $\kappa^*(A)$.
To satisfy $Atc \geq \one$, we need 
\begin{equation*}
c \geq \frac{1}{(At)_x}
\end{equation*}
for all $x$.
As long as $(At)_x > 0$ for all $x$, the vector 
\begin{equation}
z_y = \frac{t_y}{\min_{x \in X} (At)_x} \label{eq:min-deg-iter}
\end{equation}
is feasible and we have the upper bound $p^*(A) \leq \one^T z$, which we call $\kappa^*_{\textsc{mdu}}(A,t)$.
The special case
\begin{equation*}
\kappa^*_{\textsc{mdu}}(A, \one) = \frac{|Y|}{\min_{x \in X} |N_A(x)|},
\end{equation*}
is the minimum degree upper bound, which explains our choice of notation.

There is an analogous construction of a feasible point for $p^*(A)$:
\begin{equation*}
p^*_{\textsc{mdl}}(A, t) = \frac{\one^T t}{\max_{y \in Y} (A^Tt)_y}.
\end{equation*}

If a channel $A$ is input regular, then $A \one = d \one$ and 
\begin{equation*}
\kappa^*_{\textsc{mdu}}(A, \one) = \frac{|Y|}{d} = \frac{|X||Y|}{|E|},
\end{equation*}
where $E$ is the edge set of the channel graph.
If $A$ is output regular, then $\one^T A = d' \one^T$ and 
\begin{equation*}
p^*_{\textsc{mdl}}(A, \one) = \frac{|X|}{d'} = \frac{|X||Y|}{|E|}.
\end{equation*}
Consequently, if $A$ is both input and output regular, then 
\begin{equation*}
p^*(A) = \frac{|X||Y|}{|E|}.
\end{equation*}

Each bound for $p^*(A)$ or $p^*(A)$ that we present will have a vector parameter $t \in \R^Y$ or $t \in \R^X$.
Usually there is some $t$ for which the approximation is exact, but finding this $t$ is no easier that solving the original problem so this is not our motivation for including the extra parameter.
The $t$ parameter allows us to convert a single bound into an iterative approximation procedure.
Computation of the bound for some $t$ will also suggest next another $t$.
By taking this perspective for all of the bounds we consider, we can make strong comparisons between them.

\subsection{The local degree bound}
For channels that are both input and output regular, computation of the sphere packing bound $p^*$ is trivial: the minimum degree bound is exact.
However, even a single low degree input will ruin the effectiveness of the minimum degree bound.
To obtain a better upper bound on $p(A)$ and $p^*(A)$, we will construct a more sophisticated feasible point in the program for $\kappa^*(A)$ by making a small change to \eqref{eq:min-deg-iter}.
\begin{definition}
Let $A \in \mtrx{\two}{X}{Y}$ be a channel.
For $z \in \R^Y$ such that $Az > \zero$, define $\varphi_A(z) \in \R^Y$ as follows:
\begin{equation*}
\varphi_A(z)_y \deq \frac{z_y}{\min_{x \in N(y)} (Az)_x}.
\end{equation*}
Define the local degree upper bound $\kappa^*_{\textsc{ldu}}(A, z) = \one^T \varphi_A(z)$.
\end{definition}

\begin{lemma}
\label{lemma:local-iteration}
For $z \in \R^Y$ such that $z \geq \zero$ and $Az > \zero$, $\varphi_A(z)$ is feasible in the program for $\kappa^*(A)$.
If $z$ is feasible for $\kappa^*(A)$, then $\varphi_A(z) \leq z$.
\end{lemma}
\begin{IEEEproof}
To demonstrate feasibility of $\varphi(z)$, we need $\varphi(z) \geq \zero$ and $A\varphi(z) \geq \one$.
The first condition is trivially met.
For $x \in X$ and $y \in N(x)$, we have
\begin{IEEEeqnarray*}{rCl}
\varphi(z)_y &=& \frac{z_y}{\min_{t \in N(y)} (Ay)_t} \geq \frac{z_y}{(Az)_x} \\%= \frac{z_y}{\one_{N(x)}^T z}\\
(A \varphi(z))_x &=& \sum_{y \in N(x)} \varphi(z)_y \geq \frac{1}{(Az)_x} \sum_{y \in N(x)} z_y = 1
\end{IEEEeqnarray*}
and $\varphi(y)$ is feasible.

If $z$ is feasible, then $Az \geq \one$.
For all $y \in Y$ we have
\begin{IEEEeqnarray*}{rCl}
\varphi(z)_y = \frac{z_y}{\min_{x \in N(y)} (Az)_x} \leq z_y.
\end{IEEEeqnarray*}
\end{IEEEproof} 

More generally, we can view this as a single step in an iterative procedure.
Suppose that we have a vector $z \in \R^Y$ that is a feasible vector in the program for $\kappa^*(A)$.
For any channel, we can take $z = \one$ as an initial vector.
At each input $x$, the total coverage, $(Az)_x$, is at least one.
The input $x$ informs each output in $N(x)$ that it can reduce its value by a factor of $(Az)_x$.
Each output $y$ receives such a message for each input in $N(y)$, then makes the largest reduction consistent with the messages.

We can iterate this optimization step.
An iteration fails to make progress under the following condition.
From the definition $\varphi(z)_y = z_y$ if and only if $\min_{x \in N(y)} (Az)_x = 1$.
Thus $\varphi(z) = z$ if for all $y \in Y$ there is some $x \in N(y)$ such that $(Az)_x = 1$. 
This algorithm is monotonic in each entry of the feasible vector, so it cannot make progress if its input is at the frontier of the feasible space.

Scaling the input by a positive constant does not affect the output of $\varphi$: for $c \in \R$, $c>0$, $\varphi(A, z) = \varphi(A, cz)$.
We could think of $\kappa^*_{\textsc{mdl}}(A,t)$ as an involving iterative procedure as well.
It has the same scaling property.
In contrast to the local degree iteration, the maximum degree iteration always stops after a single step because the output vector is a constant multiple of the input.
The local degree iteration scales different entries in the initial vector by different amounts, so it is possible for it to make progress for multiple iterations.

\subsection{Application to the single deletion channel}
\label{section:deletion}
Let $A_n$ be the $n$-bit 1-deletion channel.
The input to the binary single deletion channel is a string $x \in [2]^n$ and the output is a substring of $x$, $y \in [2]^{n-1}$.
Each output vertex in $A_n$ has degree $n+1$.
Thus $p^*(A_n) \geq p^*_{\textsc{mdu}}(A_n) = \frac{2^n}{n+1}$.

Levenshtein \cite{levenshtein_binary_1966} showed that 
\begin{equation*}
\kappa^*(A_n) \leq \frac{2^n}{n+1}(1 + o(1)).
\end{equation*}
Kulkarni and Kiyavash computed the local degree upper bound, or equivalently $\varphi(\one)$ \cite{kulkarni_nonasymptotic_2013}.
This shows that $\kappa^*(A_n)$ is at most
\begin{equation*}
\frac{2^n}{n-1} = \frac{2^n}{n+1} \left(1 + \frac{2}{n-1}\right) = \frac{2^n}{n+1}(1 + O(n^{-1})).
\end{equation*}
Recently, Fazeli et al. found a fractional covering for $A_n$ that provides a better upper bound \cite{fazeli_generalized_2014}.
In this section, we compute $\varphi \circ \varphi(\one)$ for these channels and analyze the values of these points.
We show that Fazeli's improved covering is related to the covering $\varphi \circ \varphi(\one)$, but $\varphi \circ \varphi(\one)$ provides a better bound asymptotically.

More precisely, the upper bound from $\varphi \circ \varphi(\one)$, given in Theorem~\ref{thm:1dub}, shows that $\kappa^*(A)$ is at most
\begin{equation*}
\frac{2^n}{n-1}\left(1 - \frac{2}{n-1} + O(n^{-2})\right) = \frac{2^n}{n+1}(1 + O(n^{-2})).
\end{equation*}
The covering in Fazeli et al. gives an upper bound of
\begin{equation*}
\frac{2^n}{n+1}\left(1 + \frac{1}{n-1} + O(n^{-2})\right).
\end{equation*}

Let $r,u,b \in \N^{[2]^*}$ be vectors such that for all $x \in [2]^*$, $r_x$ is the number of runs in $x$, $u_x$ is the number of length-one runs, or unit runs, in $x$, and $b_x$ is the number of unit runs at the start or end of $x$.

Proofs of the theorems and lemmas stated in this section can be found in Appendix A.

\newcommand{\thmonetext}{Let
\begin{equation*}
f(r,u,b) \deq \frac{1}{r} \left(1 + \frac{\max (2u - b - 2,0)}{(r+2)(r+1)}\right)^{-1}.
\end{equation*}
Then the vector $z_y = f(r_y,u_y,b_y)$ is feasible for $\kappa^*(A_n)$, so $\kappa^*(A_n) \leq \one^Tz$.
}
\begin{theorem}
\label{thm:1d-feasible-pt}
\thmonetext
\end{theorem}

\begin{theorem}
\label{thm:1dub}
For $n \geq 2$,
\begin{equation*}
\kappa^*(A_n) \leq \frac{2^n}{n+1}\left(1 + \frac{26}{n(n-1)}\right).
\end{equation*}
\end{theorem}

Now we will compare this bound to the bound corresponding to the cover of Fazeli et al.
Let
\begin{equation*}
f'(r,u,b) \deq 
\begin{cases}
\frac{1}{r} \left(1 - \frac{u-b}{r^2}\right) & u - b \geq 2\\
\frac{1}{r}  & u - b \leq 1.
\end{cases}
\end{equation*}
Fazeli et al. establish that $z_y = f'(r_y,u_y,b_y)$ is feasible for $\kappa^*(A_n)$.
Compare this with the cover given by $f'$ and note that the coefficient on $u$ is 1 in $f''$ and 2 in $f'$.

\begin{lemma}
\label{lemma:feasible-pt-three}
Let $z_y = f'(r_y,u_y,b_y)$. Then 
\begin{equation*}
\one^T z \geq \frac{2^n - 2}{n+1}\left(1+\frac{1}{n-1} - \frac{3}{(n-1)(n-2)}\right)
\end{equation*}
\end{lemma}

This shows that the bound of Theorem~\ref{thm:1dub} is asymptotically better than the bound corresponding to the cover of Fazeli et al.
We could continue to iterate $\varphi$ to produce even better bounds.
The fractional covers produced would depend on more statistics of the strings.
For example, the value at a particular output of the cover produced by the third iteration of $\varphi$ would depend on the number of runs of length two in that output string, in addition to the total number of runs and the number of runs of length one.

The largest known single deletion correcting codes are the Varshamov-Tenengolts (VT) codes.
The length-$n$ VT code contains at least $\frac{2^n}{n+1}$ codewords, so these codes are asymptotically optimal.
The VT codes are known to be maximum independent sets for $n \leq 10$, but this question is open for larger $n$ \cite{sloane_challenge}.
Kulkarni and Kiyavash computed the exact value of $\kappa^*(A_n)$ for $n \leq 14$ \cite{kulkarni_nonasymptotic_2013}.
For $7 \leq n \leq 14$, the gap between $\kappa^*(A_n)$ and the size of the VT codes was at least one, so it is unlikely that sphere-packing bounds will resolve the optimality of the VT codes for larger $n$.
Despite this, it would be interesting to know whether $\kappa^*(A_n) \leq \frac{2^n}{n+1} + O(2^{cn})$ for some constant $c < 1$.

\begin{figure}
\setlength{\tabcolsep}{4pt}
\begin{tabular}{>{$}r<{$} |>{$}r<{$} >{$}r<{$} >{$}r<{$} >{$}r<{$} >{$}r<{$} >{$}r<{$} }
n & |VT_0| & p^*(A) & \text{Thm.~\ref{thm:1d-feasible-pt}} & \text{FVY} & \text{KK} & \text{Thm.~\ref{thm:1dub}} \\
\hline
 5 &        6 & 6 & 7 & 7 & 7 & 12\\
 6 &       10 & 10 & 12 & 12 & 12 & 17\\
 7 &       16 & 17 & 20 & 20 & 21 & 25\\
 8 &       30 & 30 & 35 & 35 & 36 & 41\\
 9 &       52 & 53 & 61 & 61 & 63 & 69\\
10 &       94 & 96 & 109 & 109 & 113 & 119\\
11 &      172 & 175 & 196 & 197 & 204 & 211\\
12 &      316 & 321 & 357 & 358 & 372 & 377\\
13 &      586 & 593 & 653 & 657 & 682 & 682\\
14 &     1096 & 1104 & 1205 & 1212 & 1260 & 1248\\
15 &     2048 &        & 2237 & 2251 & 2340 & 2301\\
16 &     3856 &        & 4174 & 4202 & 4368 & 4272\\
17 &     7286 &        & 7825 & 7882 & 8191 & 7977\\
18 &    13798 &        & 14727 & 14845 & 15420 & 14969\\
19 &    26216 &        & 27820 & 28059 & 29127 & 28207\\
20 &    49940 &        & 52720 & 53202 & 55188 & 53348\\
21 &    95326 &        & 100194 & 101163 & 104857 & 101226\\
22 &   182362 &        & 190912 & 192850 & 199728 & 192623\\
23 &   349536 &        & 364621 & 368478 & 381300 & 367485\\
24 &   671092 &        & 697865 & 705511 & 729444 & 702697\\
%25 &  1290556 &        & 1338261 & 1353367 & 1398101 & 1346479\\
%26 &  2485534 &        & 2570858 & 2600624 & 2684354 & 2584934\\
%27 &  4793492 &        & 4946763 & 5005268 & 5162220 & 4971026\\
%28 &  9256396 &        & 9532663 & 9647413 & 9942053 & 9574736\\
%29 & 17895736 &        & 18395355 & 18620029 & 19173961 & 18468711\\
%30 & 34636834 &        & 35543411 & 35982649 & 37025580 & 35671956\\
\end{tabular}
\caption{
The cardinality of the VT construction and several upper bounds on $p(A_n)$, where $A_n$ is the $n$-bit single deletion channel.
For $n \leq 14$, Kulkarni and Kiyavash were able computed the exact value of $p^*(A_n)$ \cite{kulkarni_nonasymptotic_2013}.
This requires solving an exponentially large linear program.
Kulkarni and Kiyavash also constructed a dual feasible point with weight $\frac{2^n-2}{n-1}$ (column KK).
This is equivalent to the first iteration of the local degree algorithm.
Fazeli et al. improved on this construction (column FVY) \cite{fazeli_generalized_2014}.
Our Theorem~\ref{thm:1d-feasible-pt} uses two interactions of the local degree algorithm.
Computing the value of the FVY and Thm.~\ref{thm:1d-feasible-pt} columns requires a sum over about $n^2$ terms.
Our Theorem~\ref{thm:1dub} gives an analytic upper bound on the weight of the feasible point from Theorem~\ref{thm:1d-feasible-pt}, which improves on existing bounds for $n \geq 22$.
}
\end{figure}

\subsection{Application to the single grain error channel}
Recently, there has been a great deal of interest in grain error channels, which are related to high-density encoding on magnetic media.
A grain in a magnetic medium has a single polarization.
If an encoder attempts to write two symbols to a single grain, only one of them will be retained.
Because the locations grain boundaries are generally unknown to the encoder, this situation can be modeled by a channel.

Mazumdar et al. applied the degree sequence bound to non-overlapping grain error channels \cite{mazumdar_coding_2011}.
Sharov and Roth applied the degree sequence bound to both non-overlapping and overlapping grain error channels \cite{sharov_bounds_2011}.
We discuss the degree sequence bound and its relationship to the local degree bound in Section~\ref{section:deg-seq}.
Kashyap and Z\'{e}mor applied the local degree bound to improve on Mazumdar et al. for the 1,2, or 3 error cases \cite{kashyap_upper_2013}.
They conjectured an extension for larger numbers of errors.
Gabrys et al. applied the local degree bound to improve on Sharov and Roth \cite{gabrys_correcting_2013}.

The input and output of this channel are strings $x,y \in [2]^n$.
To produce an output from an input, select a grain pattern with at most one grain of length two and no larger grains.
The grain of length two, if it exists, bridges indices $j$ and $j+1$ for some $0 \leq j \leq n-2$.
Then the channel output is 
\begin{equation*}
y_i = \begin{cases}
x_i & i \neq j\\
x_{i+1} & i = j
\end{cases}
\end{equation*}
If $u_j = u_{j+1}$ or if there is no grain of length two, then $y=x$.

The degree of an input string is equal to the number of runs $r$: each of the $r-1$ run boundaries could be bridged by a grain or there could be no error.
A grain error reduces the number of runs by 0,1, or 2.
The number of runs is reduced by 1 if $j=0$ and $x_0 \neq x_1$, by 2 if $j \geq 1$, $x_j \neq x_{j+1}$, and $x_{j-1} = x_{j-1}$, and by 0 otherwise.
Equivalently, the number of runs is reduced by 1 if $x$ has a length-1 run at index 0 is eliminated and by 2 if a length-1 run elsewhere is eliminated.
In the previous section, we let $u_x$ be the number of length-1 runs in $x$ and $b_x$ be the number of length-1 runs appearing at the start or end of $x$.
For the grain channel, we need to distinguish between length-1 runs at the start and at the end, so let $b_x^L$ and $b_x^R$ count these.

\begin{theorem}
\label{thm:1g-feasible-pt}
Let $A_n$ be the $n$-bit 1-grain-error channel.
The vector 
\begin{equation*}
z_y = \frac{1}{r_y} \left(1 + \frac{2u_y - 2b_y^R - b_y^L - 2}{(r_y+2)(r_y+1)}\right)^{-1}
\end{equation*}
is feasible for $\kappa^*(A_n)$.
\end{theorem}

By applying the techniques used in the proof of Theorem~\ref{thm:1dub}, it can be shown that Theorem~\ref{thm:1g-feasible-pt} implies that $\kappa^*(A_n) = \frac{2^{n+1}}{n+2}(1 + O(n^{-2}))$.

\section{The degree sequence upper bound}
\label{section:deg-seq}
The degree sequence upper bound is an important technique for dealing with nonuniform combinatorial channels that predates the local degree bound by many decades.
This is a simple generalization of the minimum degree bound and the basic idea behind the bound does not require linear programming.
For any code $C \subseteq X$, $\sum_{x \in C} |N(x)| \leq |Y|$.
The size of the largest input set $C$ satisfying this inequality is an upper bound on the size of the largest code.
This set can be found greedily by repeatedly adding the minimum degree remaining input vertex.
This approach is more robust than the minimum degree upper bound because it takes all of the input degrees into account.
Call this the degree sequence upper bound.

Levenshtein applied this idea to obtain an upper bound on codes for the deletion channel.
Kulkarni and Kiyavash applied the local degree bound to the deletion channel and showed that the resulting bound improved on Levenshtein's result.
Although its definition does not require a linear program, the degree sequence bound still has a nice linear programming interpretation.
Taking this perspective, we compare the performance of the degree sequence bound to the local degree bound for arbitrary channels and show that the local degree bound is always better.
The degree sequence bound uses less information about the channel than the local degree bound and achieves a weaker but still not trivial result.
At the end of this section, we discuss bounds that use even less information the degree sequence bound.

\subsection{Linear programs for the degree sequence bound}
While the local degree upper bound is naturally expressed as a feasible point in the program for $\kappa^*$, the easiest linear programming interpretation of the degree sequence bound works differently.
The degree sequence upper bound is the value of a further relaxation of the program for $p^*$.
It turns out that the minimum degree upper bound is easily expressed as both a dual feasible point and a primal relaxation.
Section~\ref{section:iterative} included the former interpretation and the latter is given here.
For a channel $A \in \mtrx{\two}{X}{Y}$ and a vector $t \in \R^Y$, define
\lpp{p^*_{\textsc{mdu}}(A,t)}{=}{\max_{w \in \R^X}}{\one^T w}{w \geq \zero}{t^TA^Tw \leq t^T \one.}
The solution to this program puts all of the weight on the minimum degree input $\argmin_x (At)_x$, so $p^*_{\textsc{mdu}}(A,t) = \kappa^*_{\textsc{mdu}}(A,t)$.

Recall that $A \one$ is the vector of input degrees of the channel graph of $A$.
Thus the main constraint of the program for $\kappa^*_{\textsc{mdu}}(A, \one)$ is $\sum_{x \in X} |N(x)| w_x \leq |Y|$.
In a code, each vertex can only be included once.
We can capture this fact and improve the upper bound by adding the additional constraint $w \leq \one$ to the program.

\begin{definition}
For a channel $A \in \mtrx{\two}{X}{Y}$ and a vector $t \in \R^Y$, define the degree sequence bound
\lpp{p^*_{\textsc{dsu}}(A,t)}{=}{\max_{w \in \R^X}}{\one^T w}{\zero \leq w \leq \one}{t^TA^Tw \leq t^T \one.}
\end{definition}

The degree sequence upper bound is tight: for a given input degree distribution and output space size, there is some channel where the neighborhoods of the small degree inputs are disjoint.
For this channel, the degree sequence upper bound is tight.
The bound cannot be improved with incorporating more information about the structure of the channel.

The local degree upper bound, which incorporates information about the channel beyond the degree sequence, is always at least as good as the degree sequence bound.
To do this, we associate the degree sequence bound with a particular feasible point in the program for $\kappa^*(A)$.

\begin{theorem}
\label{thm:dt-vs-ld}
Let $A \in \mtrx{\two}{X}{Y}$ be a channel and let $t \in \R^Y$.
Let $d \in \R$ be a degree threshold and define the following sets of inputs:
\begin{IEEEeqnarray*}{rCl}
X_- &=& \{x \in X : (At)_x < d\},\\
X_0 &=& \{x \in X : (At)_x = d\}.
\end{IEEEeqnarray*}
If $\one_{X_-}^TAt \leq \one^T t \leq (\one_{X_-}^T + \one_{X_0}^T)At$, then the point 
\begin{equation*}
z_y = t_y\left( \frac{1}{d} + \sum_{x \in N(y)} \max \left(\frac{1}{(At)_x} - \frac{1}{d}, 0 \right) \right)
\end{equation*}
is feasible in $\kappa^*(A)$, $p^*_{\textsc{dsu}}(A, t) = \one^T z$, and $\varphi_A(t) \leq z$.
Thus $\kappa^*_{\textsc{ldu}}(A,t) \leq p^*_{\textsc{dsu}}(A,t)$.
\end{theorem}
\begin{IEEEproof}
First we establish that $p^*_{\textsc{dsu}}(A,t) = |X_-| + \frac{\one^T t - \one_{X_-}^TAt}{d}$ by constructing a primal feasible point and a dual feasible point with this value.

The point $w = \one_{X_-} + \frac{\one^T t - \one_{X_-}^TAt}{d|X_0|}\one_{X_0}$ is feasible for the primal program.
This puts the maximum possible weight on each of the inputs with degree below the threshold and fractional weight on inputs with degree equal to the threshold.

The dual program is 
\lpp{}{}{\min_{z'_0 \in \R,\,z' \in \R^X}}{\one^T t z'_0 + \one^T z'}{(z'_0,z') \geq \zero}{A t z'_0 + z' \geq \one.}
The point $z'_0 = \frac{1}{d}$, $z'_x = \max(0,\frac{d-(At)_x}{d})$ is feasible in the dual program.
Note that $z'_x > 0$ only for $x \in X_-$.
The value of this point is 
\begin{equation*}
\frac{\one^T t}{d} + \sum_{x \in X_-} \frac{d - (At)_x}{d} = |X_-| + \frac{\one^T t - \one_{X_-}^TAt}{d}.
\end{equation*}

We construct $z$ from $(z'_0,z')$.
Let 
\begin{equation*}
z = \diag{t} (\one z'_0 + A^T \diag{At}^{-1} z')
\end{equation*}
Then 
\begin{IEEEeqnarray*}{rCl}
\one^T z' 
&=& \one^T \diag{t}(\one z'_0 + A^T \diag{At}^{-1} z') \\
&=& t^T \one z'_0 + t^TA^T \diag{At}^{-1} z' \\
&=& \one^T t z'_0 + \one^T z' 
\end{IEEEeqnarray*}
so $z$ has the necessary weight.
Substituting the values $z_0 = \frac{1}{d}$ and $z_x = \max(\frac{d-(At)_x}{d}, 0)$, we obtain
\begin{IEEEeqnarray*}{rCl}
\frac{z'_y}{t_y} &=& \frac{1}{d} + \sum_{x \in N(y)} \frac{1}{(At)_x} \max \left( \frac{d - (At)_x}{d}, 0 \right) \\
&=& \frac{1}{d} + \sum_{x \in N(y)} \max \left(\frac{1}{(At)_x} - \frac{1}{d}, 0 \right)\\
&\geq& \frac{1}{d} + \max_{x \in N(y)} \max \left(\frac{1}{(At)_x} - \frac{1}{d}, 0 \right)\\
&=& \max_{x \in N(y)} \max \left(\frac{1}{(At)_x}, \frac{1}{d} \right)\\
&\geq& \max_{x \in N(y)} \frac{1}{(At)_x}\\
&=& \frac{1}{t_y} \varphi_A(t).
\end{IEEEeqnarray*}
Because $\varphi_A(t)$ is feasible in $\kappa^*(A)$, $z$ is as well.
\end{IEEEproof}
This interpretation of the degree sequence bound allows us to iteratively construct dual feasible points, but these points are dominated by those produced by the local degree algorithm.
However, this interpretation does give some intuition about the source of the superior performance of the local degree bound.
When multiple low-degree inputs have a common output, the local degree bound takes advantage of this fact.
This information is not contained in the degree distribution.

The degree sequence upper bound is monotonic and decreasing in the degree of each vertex.
For a complicated channel, as a tractable alternative to computing the exact degree sequence bound, we can compute the bound for a simplified degree sequence.
Pick a degree threshold $d$.
Treat each vertex with degree at least $d$ as if it had degree exactly $d$ and treat each vertex with degree less than $d$ as if it had degree one.
The bound coming from this modified degree sequence is
\begin{equation*}
p^*_{\textsc{dsu}}(A,\one) \leq \frac{|X| - |S|}{d} + |S|,
\end{equation*}
where $S = \{x \in X : |N(x)| < d\}$, the set of low degree inputs.
This approach is effective when the degree distribution concentrates around its mean but still has a few vertices with much lower degree.

\subsection{Bounds that use only the number of edges}
\label{subsection:edge-bounds}
The degree sequence bound uses the full input degree distribution of the channel graph and the local degree bounds use the degrees of the endpoints of each edge.
Suppose that we only know the number of inputs, output, and edges in the channel graph.
This means that we know the average input degree and the average output degree but nothing else about the degree distributions.
Recall from Section~\ref{section:iterative} that $\kappa^*(A) = \frac{|X||Y|}{|E|}$ for a channel $A \in \mtrx{\two}{X}{Y}$ with constant input and output degrees.
%The average (weighted) input degree of a channel $A$ is $\frac{\one^T A t}{|X|}$.
%The point $\frac{\one^T t}{\one^T A t} \one \in \R^X$ is feasible in the primal program for $p^*_{\textsc{dsu}}(A,t)$, so $\frac{|X| \one^T t}{\one^T A t} \leq p^*_{\textsc{dsu}}(A,t)$.
The point $\frac{|Y|}{|E|} \one \in \R^X$ is feasible in the primal program for $p^*_{\textsc{dsu}}(A,\one)$, so $\frac{|X||Y|}{|E|} \leq p^*_{\textsc{dsu}}(A,\one)$.
To give us a relationship between $\kappa^*(A)$ and the average input degree of $A$, we would need an inequality running in the other direction.
It turns out that the number of edges in a bipartite graph gives us weak bounds on the packing and covering numbers for the graph.

\begin{lemma}
\label{lemma:edge-only-ub}
Let $A \in \mtrx{\two}{X}{Y}$ be a channel and let $E$ be the edge set of the channel graph. Then
\begin{equation}
\kappa(A) \leq |X| - \frac{|E|}{|Y|} + 1. \label{eq:eoub}
\end{equation}
For any $X$, $Y$, and $S \subseteq X$ such that $|S| \leq |Y|$, there is a channel $A$ such that $|E| =|Y|(|X|-|S|+1)$ and $S$ is an input packing in $A$.
Thus \eqref{eq:eoub} cannot be improved.
\end{lemma}\begin{IEEEproof}
For any output $y \in Y$, we can construct a cover using $y$ together with $|X| - |N(y)|$ other outputs: for each $x \in X \setminus N(y)$, we add an arbitrary member of $N(x)$ to our cover.
Because $\sum_{y \in Y} |N(y)| = |E|$, there is some $y$ with $|N(y)| \geq \frac{|E|}{|Y|}$.

We construct the tightness example $A$ as follows.
Choose the neighborhoods of the inputs in $S$ so that each is nonempty, they are disjoint, and $\bigcup_{x \in S} N(x) = Y$.
Meeting the first two conditions is possible is possible because $|S| \leq |Y|$.
Because the neighborhoods are disjoint, $S$ is a packing. 
For each $x \in X \setminus S$, let $N(x) = Y$.
Then all output degrees are all equal to $|X|-|S|+1$.
\end{IEEEproof}
In other words, a large number of edges are necessary to rule out the existence of a set of input vertices with small degree.

\section{Confusability graphs and families of channels}
Sphere packing upper bounds are obtained from combinatorial channels.
However, for any channel there is a simpler object that also characterizes the set of codes: the confusability graph.
Furthermore, any particular confusability graph arises from many combinatorial channels.
To obtain upper bounds on the size of codes for one channel it can be useful to consider the sphere packing bounds that arise from some other equivalent channel.
At the end of this section, we show how the Hamming and Singleton bounds are an example of this phenomenon.

\subsection{Confusability graphs and independent sets}
\label{subsection:confusion-graph}
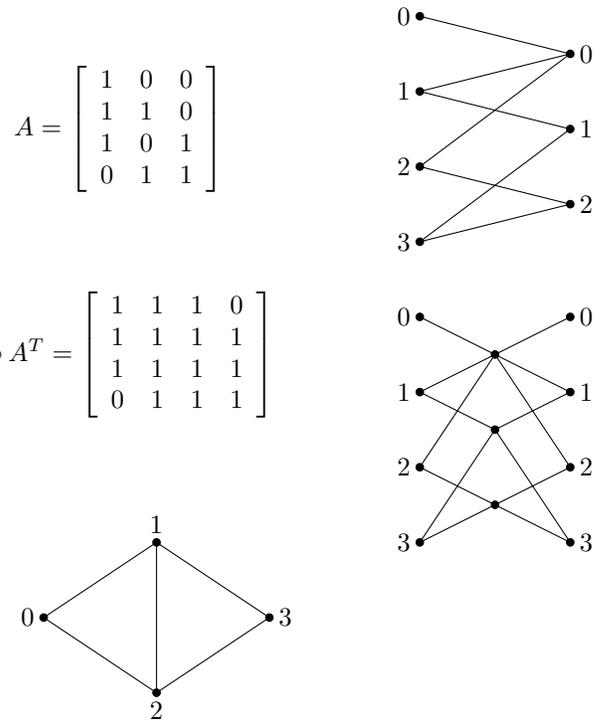
\begin{figure}
 \begin{tikzpicture}[auto]

   \node (mA)  at (0, 5.5) {$A = 
     \left[ \begin{array}{ccc} 
         1 & 0 & 0\\
         1 & 1 & 0\\
         1 & 0 & 1\\
         0 & 1 & 1\\
       \end{array} \right]$};

   \node (mAoA)  at (0, 2.5) {$A \circ A^T = 
     \left[ \begin{array}{cccc} 
         1 & 1 & 1 & 0\\
         1 & 1 & 1 & 1\\
         1 & 1 & 1 & 1\\
         0 & 1 & 1 & 1\\
       \end{array} \right]$};

   \coordinate[label=left:$0$] (i0) at (4,7);
   \coordinate[label=left:$1$] (i1) at (4,6);
   \coordinate[label=left:$2$] (i2) at (4,5);
   \coordinate[label=left:$3$] (i3) at (4,4);

   \coordinate[label=right:$0$] (o0) at (6,6.5);
   \coordinate[label=right:$1$] (o1) at (6,5.5);
   \coordinate[label=right:$2$] (o2) at (6,4.5);

   \draw (i0) -- (o0);
   \draw (i1) -- (o0) -- (i2) -- (o2) -- (i3) -- (o1) -- (i1);

   \draw[fill] (i0) circle [radius=0.05];
   \draw[fill] (i1) circle [radius=0.05];
   \draw[fill] (i2) circle [radius=0.05];
   \draw[fill] (i3) circle [radius=0.05];
   \draw[fill] (o0) circle [radius=0.05];
   \draw[fill] (o1) circle [radius=0.05];
   \draw[fill] (o2) circle [radius=0.05];

   \coordinate[label=left:$0$] (j0) at (4,3);
   \coordinate[label=left:$1$] (j1) at (4,2);
   \coordinate[label=left:$2$] (j2) at (4,1);
   \coordinate[label=left:$3$] (j3) at (4,0);

   \coordinate (k0) at (5,2.5);
   \coordinate (k1) at (5,1.5);
   \coordinate (k2) at (5,0.5);

   \coordinate[label=right:$0$] (l0) at (6,3);
   \coordinate[label=right:$1$] (l1) at (6,2);
   \coordinate[label=right:$2$] (l2) at (6,1);
   \coordinate[label=right:$3$] (l3) at (6,0);

   \draw (j0) -- (k0) -- (l0);
   \draw (j1) -- (k0) -- (j2) -- (k2) -- (j3) -- (k1) -- (j1);
   \draw (l1) -- (k0) -- (l2) -- (k2) -- (l3) -- (k1) -- (l1);

   \draw[fill] (j0) circle [radius=0.05];
   \draw[fill] (j1) circle [radius=0.05];
   \draw[fill] (j2) circle [radius=0.05];
   \draw[fill] (j3) circle [radius=0.05];
   \draw[fill] (k0) circle [radius=0.05];
   \draw[fill] (k1) circle [radius=0.05];
   \draw[fill] (k2) circle [radius=0.05];
   \draw[fill] (l0) circle [radius=0.05];
   \draw[fill] (l1) circle [radius=0.05];
   \draw[fill] (l2) circle [radius=0.05];
   \draw[fill] (l3) circle [radius=0.05];

   \coordinate[label=left:$0$] (g0) at (-1,-1);
   \coordinate[label=above:$1$] (g1) at (0.5,0);
   \coordinate[label=below:$2$] (g2) at (0.5,-2);
   \coordinate[label=right:$3$] (g3) at (2,-1);

   \draw (g0) -- (g1) -- (g3) -- (g2) -- (g0);
   \draw (g1) -- (g2);
   \draw[fill] (g0) circle [radius=0.05];
   \draw[fill] (g1) circle [radius=0.05];
   \draw[fill] (g2) circle [radius=0.05];
   \draw[fill] (g3) circle [radius=0.05];
 \end{tikzpicture}
\caption{
At the top, a channel $A \in \mtrx{\two}{[4]}{[3]}$ with its corresponding channel graph.
In the middle, the channel $A \circ A^T$, and the composition of the channel graph of $A$ with the channel graph of $A^T$.
$(A \circ A^T)_{i,j} = 1$ if there is a path going left to right from $i$ to $j$ in that graph.
At the bottom, the confusability graph of $A$, which has adjacency matrix $A \circ A^T - I$.
}
\label{fig:channel}
\end{figure}

\begin{definition}
Let $A \in \mtrx{\two}{X}{Y}$ and $B \in \mtrx{\two}{Y}{Z}$ be channels.
Then define $A \circ B \in \mtrx{\two}{X}{Z}$, the composition of $A$ and $B$, such that
\begin{equation*}
N_{A \circ B}(x) = \bigcup_{y \in N_A(x)} N_B(y)
\end{equation*}
\end{definition}
We can characterize $A \circ B$ in two other ways:
\begin{IEEEeqnarray}{rCl}
(A \circ B)_{x,z} &=& 
\begin{cases}
1 & N_A(x) \cap N_B(z) \neq \varnothing \label{eq:intersection}\\
0 & N_A(x) \cap N_B(z) = \varnothing
\end{cases}\\
 &=& \max_{y \in Y} \min(A_{x,y}, B_{y,z}) \label{eq:semiring}
\end{IEEEeqnarray}
The characterization \eqref{eq:semiring} states that $A \circ B$ is the matrix product of $A$ and $B$ in the Boolean semiring.

Let $I$ denote the identity matrix.
\begin{definition}
\label{def:confusion}
For a channel $A \in \mtrx{\two}{X}{Y}$, define the confusability graph of $A$ to be the graph with vertex set $X$ and adjacency matrix $A \circ A^T - I$.
\end{definition}
Because $A \circ A^T - I$ is a zero-one symmetric matrix with zeros on the diagonal, the confusability graph is simple and undirected.
From \eqref{eq:intersection}, vertices $u$ and $v$ are adjacent in the confusability graph of $A$ if and only if $N(u)$ and $N(v)$ intersect.
Figure~\ref{fig:channel} shows an example of a channel, its composition with its reverse, and its confusability graph.

\begin{definition}
Let $G$ be an undirected simple graph with vertex set $X$.
A set $S \subseteq X$ is independent in $G$ if and only if for all $u, v \in S$, $u$ and $v$ are not adjacent.
The maximum size of an independent set in $G$ is denoted by $\alpha(G)$.
\end{definition}
Now we have a second important characterization of codes.
\begin{lemma}
\label{lemma:ind-p}
Let $G$ be the confusability graph for a channel $A \in \mtrx{\two}{X}{Y}$.
Then a set $C \subseteq X$ is code for a $A$ if and only if it is an independent set in $G$.
Thus $\alpha(G) = p(A)$.
\end{lemma}
\begin{IEEEproof}
A set $C$ is not a code if and only if there is some $y$ such that $N(y)$ contains distinct codewords $u$ and $v$, or equivalently $y \in N(u) \cap N(v)$.
This means $(A \circ A^T)_{u,v} = 1$, $u$ and $v$ are adjacent in the confusability graph, and $C$ is not independent.
\end{IEEEproof}
The confusability graph does not contain enough information to recover the original channel graph, but it contains enough information to determine whether a set is a code for the original channel.

\subsection{Families of channels with the same codes}
\label{section:families}
There are many different channels that have $G$ as a confusability graph.
A \emph{clique} in a graph $G$ is a set of vertices $S$ such that for all distinct $u,v \in S$, $\{u,v\} \in E(G)$.
If $G$ is the confusability graph for a channel $A \in \mtrx{\two}{X}{Y}$, then for each $y \in Y$, $N(y)$ is a clique in $G$.
Let $\Omega \subseteq 2^X$ be a family of cliques that covers every edge in $G$.
This means that for all $\{u,v\} \in E(G)$, there is some $S \in \Omega$ such that $u,v \in S$.
Let $H \in \mtrx{\two}{X}{\Omega}$ be the vertex-clique incidence matrix: $H_{x,S} = 1$ is $x \in S$ and $H_{x,S} = 0$ otherwise.
Then $\alpha(G) = p(H)$.

Thus each family of cliques that covers every edge gives us an integer linear program that expresses the maximum independent set problem for $G$.
These programs all contain the same integer points, the indicators of the independent sets of $G$.
However, their polytopes are significantly different so the fractional relaxations of these programs give widely varying upper bounds on $\alpha(G)$.

Each edge in $G$ is a clique, so $E(G)$ is one natural choice for $\Omega$.
Then $\alpha(G) = p(H_E)$, where $H_E \in \mtrx{\two}{X}{E(G)}$ is the vertex edge incidence matrix for $G$.
However, relaxing the integrality constraint for this program gives a useless upper bound.
The vector $w = \frac{1}{2} \one$ is feasible, so $p^*(H_E) \geq \frac{|X|}{2}$ regardless of the structure of $G$.
%The matrix $A$ is easy to construct from any representation of $G$.

\begin{definition}
Let $\Omega$ be the set of \emph{maximal} cliques in $G$ and let $H_{\Omega} \in \mtrx{\two}{X}{\Omega}$ be the vertex-clique incidence matrix.
Then $\alpha(G) = p(H_{\Omega})$.
Define the minimum clique cover of $G$, $\theta(G) \deq \kappa(H_{\Omega})$ and the minimum fraction clique cover $\theta^*(G) \deq \kappa^*(H_{\Omega})$.
\end{definition}
Unlike the program derived from the edge set, $\theta^*(G)$ gives a nontrivial upper bound on $\alpha(G)$.
In fact, $\theta^*(G)$ is the best sphere packing bound for any channel that has $G$ as its confusability graph.

\begin{lemma}
\label{lemma:clique-edge-cover-dominance}
Let $G$ be a graph with vertex set $X$ and let $\Omega_1,\Omega_2 \subseteq 2^X$ be families of cliques that cover every edge in $G$.
Let $H_1, H_2$ be the vertex-clique incidence matrices for $\Omega_1$ and $\Omega_2$ respectively.
If for each $R \in \Omega_1$ there is some $S \in \Omega_2$ such that $R \subseteq S$, then $p^*(H_2) \leq p^*(H_1)$. 
\end{lemma}
\begin{IEEEproof}
A clique $S$ gives the constraint $\sum_{x \in R} w_x \leq 1$ in $p$.
If $R \in \Omega_1$, $S \in \Omega_2$, and $R \subseteq S$, then the constraint from $R$ is implied by the constraint for $S$.
Any additional cliques in $\Omega_2$ can only reduce the feasible space for $p(H_2)$.
Thus the feasible space for $p(H_2)$ is contained in the feasible space for $p(H_1)$.
\end{IEEEproof}
\begin{corollary}
\label{corr:theta}
Let $A \in \mtrx{\two}{X}{Y}$ be a channel and let $G$ be the confusability graph for $A$. Then $\theta^*(G) \leq \kappa^*(A)$.
\end{corollary}
\begin{IEEEproof}
For each output $y \in Y$, $N(y)$ is a clique in $G$ and these clique cover every edge of $G$.
Each clique in $G$ is contained in a maximal clique, so the claim follows immediately from Lemma~\ref{lemma:clique-edge-cover-dominance}.
\end{IEEEproof}

Corollary~\ref{corr:theta} suggests that we should ignore the structure of our original channel $A$ and try to compute $\theta^*(G)$ instead of $\kappa^*(A)$.
However, there is no guarantee that we can efficiently construct the linear program for $\theta^*(G)$ by starting with $G$ and searching for all of the maximal cliques.
We are often interested in graphs with an exponential number of vertices.
Even worse, the number of maximal cliques in $G$ can exponentially in the number of vertices.
To demonstrate this, consider a complete $k$-partite graph with 2 vertices in each part.
%In this graph, each vertex is adjacent to all vertices except those in its part.
If we select one vertex from each part, we obtain a maximal (and also maximum) clique.
The graph has $2k$ vertices, but there are $2^k$ maximal cliques.

The fractional clique cover number has been considered in the coding theory literature in connection with the Shannon capacity of a graph, $\Theta(G)$.
The capacity of a combinatorial channel $A$ is $\lim_{n \rightarrow \infty} p(A^n)^{\frac{1}{n}}$, the number of possible messages per channel use when the channel can be used many times.
Like $p(A)$, the capacity of the channel depends only on its confusability graph.
Thus the Shannon capacity of a graph $G$ can be defined as the capacity of a channel with confusability graph $G$.
The Shannon capacity of a graph is at least as large as the maximum independent set and is extremely difficult to compute.
Shannon used something equivalent to a clique cover as an upper bound for Shannon capacity \cite{shannon_zero_1956}.
Rosenfeld showed the connection between Shannon's bound and linear programming \cite{rosenfeld_problem_1967}.
Shannon also showed that the feedback capacity of a combinatorial channel $A$ is $p^*(A)$.
Lovasz introduced the Lovasz theta function of a graph, $\vartheta(G)$, and showed that it was always between the Shannon capacity and the fractional clique cover number \cite{lovasz_shannon_1979}.
All together, we have
\begin{equation*}
\alpha(G) \leq \Theta(G) \leq \vartheta(G) \leq \theta^*(G).
\end{equation*}
The Lovasz theta function is derived via semidefinite programming and consequently is not a sphere-packing bound.

There are also several connections between these concepts and communication over probabilistic channels.
For a combinatorial channel $A$, the minimum capacity over the probabilistic channels with support $A$ is $p^*(A)$.
Recently Dalai has proven upper bounds on the reliability function of a probabilistic channel that are finite for all rates above at the (logarithmic) Shannon capacity of the underlying confusion graph, in contrast to previous bounds that were finite for rates above $\log p^*(A)$ \cite{dalai_lower_2013}.
The idea of multiple channels with the same confusion graph plays an important role here.

\subsection{Hamming and Singleton Bounds}
Sometimes a channel has some special structure that allows us to find an easily described family of channels with the same codes.
Then we can optimize over the family by computing the a bound for each channel and using the best.
This technique has been successfully applied to deletion-insertion channels by Cullina and Kiyavash\cite{cullina_improvement_2013}.
Any code capable of correcting $s$ deletions can also correct any combination of $s$ total insertions and deletions.
Two input strings can appear in an $s$-deletion-correcting code if and only if the deletion distance between them is more than $s$.
In the asymptotic regime with $n$ going to infinity and $s$ fixed, each channel in the family becomes approximately regular.
Thus the degree threshold bound gives a good approximation to the exact sphere-packing bound for these channels.
The best bound comes from a channel that performs approximately $\frac{qs}{q+1}$ deletions and $\frac{s}{q+1}$ insertions, where $q$ is the alphabet size.

In this section we present a very simple application of this technique.
Consider the channel that takes a $q$-ary vector of length $n$ as its input, erases $a$ symbols, and substitutes up to $b$ symbols.
Thus there are $q^n$ channel inputs, $\binom{n}{a}q^{n-a}$ outputs, and each input can produce $\binom{n}{a}\sum_{i=0}^b \binom{n-a}{i}(q-1)^i$ possible outputs.
Two inputs share a common output if and only if their Hamming distance is at most $s=a+2b$.
For each choice of $n$ and $s$, we have a family of channels with identical confusability graphs.
Call the $q$-ary $n$-symbol $a$-erasure $b$-substitution channel $A_{q,n,a,b}$.
These channels are all input and output regular, so 
\begin{IEEEeqnarray*}{rCl}
\kappa^*(A_{q,n,a,b}) &=& \frac{\binom{n}{a}q^{n-a}}{\binom{n}{a}\sum_{i=0}^b \binom{n-a}{i}(q-1)^i}\\
&=& \frac{q^{n-a}}{\sum_{i=0}^b \binom{n-a}{i}(q-1)^i}.
\end{IEEEeqnarray*}

Two special cases give familiar bounds.
For even $s$, setting $a=0$ and $b=s/2$ produces the Hamming bound:
\begin{equation*}
\kappa^*(A_{q,n,0,s/2}) = \frac{q^n}{\sum_{i=0}^{s/2} \binom{n}{i}(q-1)^i}.
\end{equation*}
Setting $a=s$ and $b=0$ produces the Singleton bound:
\begin{equation*}
\kappa^*(A_{q,n,s,0}) = q^{n-s}.
\end{equation*}

For $q=2$, the Hamming bound is always the best bound in this family.
When $q$ is at least 3, each bound in the family is the best for some region of the parameter space.

\begin{lemma}
\label{lemma:threshold}
$\kappa^*(A_{q,n,a,b}) \leq \kappa^*(A_{q,n,a+2,b-1})$ when $a + qb \leq n-1$.
\end{lemma}
The proof of Lemma~\ref{lemma:threshold} can be found in Appendix A.
\begin{theorem}
Let $q,n,s \in \N$ such that $q \geq 3$, $0 \leq s \leq n-1$, and $s$ even.
Then 
\begin{IEEEeqnarray*}{rCl}
\argmin_{0 \leq b \leq s/2} \kappa^*(A_{q,n,s-2b,b})
&=&\begin{cases}
s/2                                         & s \leq \frac{2}{q}(n-1) \\
\left\lfloor \frac{n-1-s}{q-2}\right\rfloor & s \geq \frac{2}{q}(n-1)
\end{cases}
\end{IEEEeqnarray*}
For fixed $\delta$,$\frac{2}{q} \leq \delta \leq 1$, and $s = \delta n$ 
\begin{equation*}
\lim_{n \rightarrow \infty} \frac{1}{n} \log \min_{0 \leq b \leq s/2} \kappa^*(A_{q,n,s-2b,b}) = (1-\delta) \log(q-1).
\end{equation*}
\end{theorem}
\begin{IEEEproof}
Let $a+2b=s$, so $a+qb = s + (q-2)b$.
From Lemma~\ref{lemma:threshold}, $\kappa^*(A_{q,n,0,s/2})$ is the smallest in the family when $s + (q-2)\frac{s}{2} \leq n-1$ or equivalently $s \leq \frac{2}{q}(n-1)$.

For $b \geq 1$ the following are equivalent:
\begin{IEEEeqnarray*}{rCcCl}
\kappa^*(A_{q,n,a+2,b-1}) &\geq& \kappa^*(A_{q,n,a,b}) &\leq& \kappa^*(A_{q,n,a-2,b+1})\\
s + (q-2)b &\leq& n-1 &\leq& s + (q-2)(b+1)\\
b &\leq& \frac{n-1-s}{q-2} &\leq& b+1.
\end{IEEEeqnarray*}
Let $b^*$ be the optimal choice of $b$.
Then 
\begin{IEEEeqnarray*}{rcCl}
\lim_{n \rightarrow \infty} & \frac{b^*}{n} &=& \frac{1 - \delta}{q-2},\\ 
\lim_{n \rightarrow \infty} & \frac{n-s+2b^*}{n} &=& 1 - \delta + 2\frac{1 - \delta}{q-2} = \frac{q(1 - \delta)}{q-2},\\
\lim_{n \rightarrow \infty} & \frac{b^*}{n-s+2b^*} &=& \frac{1}{q}.
\end{IEEEeqnarray*}
Finally,
\begin{IEEEeqnarray*}{Cl}
&\lim_{n \rightarrow \infty} \frac{1}{n} \log \frac{q^{n - s + 2b^*}}{\sum_{i=0}^{b^*} \binom{n-s+2b^*}{i} (q-1)^i}\\
= & \lim_{n \rightarrow \infty} \frac{n - s + 2b^*}{n} \log q\\
& - \frac{n-s+2b^*}{n} H_2 \left(\frac{b^*}{n-s+2b^*}\right) - \frac{b^*}{n} \log (q-1)\\
= & \frac{q(1-\delta)}{q-2} \log q - \frac{q(1-\delta)}{q-2}H_2(1/q) - \frac{1-\delta}{q-2} \log (q-1)\\
= & \frac{1-\delta}{q-2}\left(q \log q - \log q - (q-1)\log \frac{q}{q-1} - \log (q-1)\right)\\
= & \frac{1 - \delta}{q-2}((q-1) \log (q-1) - \log (q-1))\\
= & (1 - \delta) \log (q-1)
\end{IEEEeqnarray*}
which proves the last claim.
\end{IEEEproof}

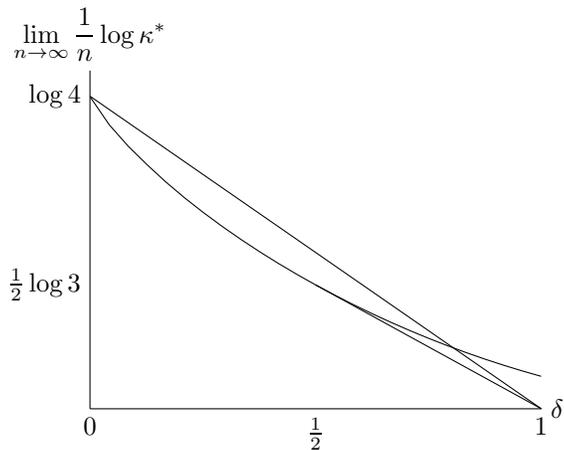
\begin{figure}
\begin{tikzpicture}[domain=0:1, xscale=6, yscale=3]
\draw (0,0) -- (1,0) node[right] {$\delta$};
\draw (0,0) -- (0,1.5) node[above] {$\displaystyle \lim_{n \rightarrow \infty} \frac{1}{n} \log \kappa^*$};
\draw (0,0) node[below] {$0$};
\draw (0.5,0) node[below] {$\frac{1}{2}$};
\draw (1,0) node[below] {$1$};
\draw (0,{ln(4)}) node[left] {$\log 4$};
\draw (0,{ln(3)/2}) node[left] {$\frac{1}{2}\log 3$};

\draw plot[id=singleton] (\x,{ln(4) * (1-\x)});
\draw plot[id=hamming,domain=0.001:1]   (\x,{ln(2) + (\x * ln(\x) + (2-\x) * ln(2-\x) - \x * ln(3)) / 2});
\draw plot[id=mixed,domain=0.5:1] (\x,{ln(3)*(1-\x)});
\end{tikzpicture}
\caption{
Sphere-packing bounds for channel performing substitution errors and erasures.
The curved line is the Hamming bound, which is $\lim_{n \rightarrow \infty} \frac{1}{n} \log\kappa^*(A_{4,n,0,s/2})$.
The upper straight line is the Singleton bound, which is $\lim_{n \rightarrow \infty} \frac{1}{n} \log \kappa^*(A_{4,n,s,0})$.
The straight line running from $(\frac{1}{2}, \frac{1}{2}\log 3)$ to $(1,0)$ is the optimized sphere-packing bound, $\lim_{n \rightarrow \infty} \frac{1}{n} \log \min_{0 \leq b \leq s/2} \kappa^*(A_{4,n,s-2b,b})$.
}
\label{fig:hamming}
\end{figure}

This family of bounds fills in the convex hull of the Hamming and Singleton bounds.
Figure~\ref{fig:hamming} plots this optimized bound, the Hamming bound, and Singleton bound for $q=4$.

There are several open questions regarding families of channels with the same confusability graphs.
Under what conditions can we find these families?
What is the relationship between these families and distance metrics?
When we have a family of channels that are not input or output regular, what should we do to get the best bounds?

\section{Lower bounds}
\label{section:lower-bounds}
In this section, we will discuss two sources of lower bounds on the size of an optimal code: sphere-covering bounds related to the dominating set problem and bounds related to Tur\'{a}n's theorem.
The dominating set problem is closely connected to linear programming, so there is a local degree sphere-covering lower bound.
This lower bound has some interesting relationships with the Tur\'{a}n type bounds.

\subsection{Sphere-covering and dominating sets}
\label{subsection:covering}
Recall a few facts from Section~\ref{subsection:confusion-graph}.
Codes for a channel are independent sets in the confusability graph for that channel.
For a channel $A \in \mtrx{\two}{X}{Y}$, the confusability graph $G$ has adjacency matrix $B - I$ where $B = A \circ A$.
The size of the largest code is $\alpha(G) = p(A)$.
We will use these notational conventions throughout this section.

\begin{definition}
Let $G$ be a graph with vertex set $X$.
A set $S \subseteq X$ is dominating in $G$ if and only if for all $x \in X \setminus S$, there is some $u \in S$, such that $x$ and $u$ are adjacent.
The minimum size of a dominating set in $G$ is denoted by $\gamma(G)$.
\end{definition}
Dominating set is a covering problem.
A vertex $u \in S$ covers itself and all adjacent vertices.
Let $G$ be a simple graph with vertex set $X$ and adjacency matrix $B-I$.
Then $S \subseteq X$ is a dominating set in $G$ if and only if $S$ is an output covering for $B$.
Thus $\gamma(G) = \kappa(B)$.

We are interested in dominating sets because of their relationship with independent sets.
\begin{lemma}
\label{lemma:dom-ind}
For any graph $G$, $\gamma(G) \leq \alpha(G)$.
\end{lemma}
\begin{IEEEproof}
If no additional vertices can be added to an independent set, each vertex of $G$ is either in the independent set or adjacent to a vertex in the independent set.
Consequently, any maximal independent set is dominating.
\end{IEEEproof}

Because dominating set is a minimization problem, its fractional relaxation is a lower bound.
Overall, we have
\begin{equation*}
p^*(B) = \kappa^*(B) \leq \kappa(B) = \gamma(G) \leq \alpha(G) = p(A).
\end{equation*}
This motivates simple lower bounds for $p^*(B)$.
In Sections~\ref{section:iterative} and \ref{section:deg-seq}, we discussed the minimum degree, degree sequence, and local degree upper bounds.
Each of these has a lower bound analogue.
The maximum degree lower bound is
\begin{IEEEeqnarray*}{rCl}
p^*_{\textsc{mdl}}(A, t) = \frac{\one^T t}{\max_{y \in Y} (A^Tt)_y}.
\end{IEEEeqnarray*}
The degree sequence lower bound is
\lpp{p^*_{\textsc{dsu}}(A,t)}{=}{\max_{w \in \R^X}}{\one^T w}{\zero \leq w \leq \one}{t^TA^Tw \leq t^T \one.}
The local degree lower bound is
\begin{IEEEeqnarray*}{rCl}
\kappa^*_{\textsc{ldl}}(A, t) = \sum_{x \in X} \frac{t_x}{\max_{y \in N(x)} (A^Tt)_y}.
\end{IEEEeqnarray*}
These satisfy the same inequalities as the upper bound versions but in the reverse order:
\begin{equation*}
p^*_{\textsc{mdl}}(A, t) \leq p^*_{\textsc{dsl}}(A, t) \leq \kappa^*_{\textsc{ldl}}(A, t) \leq p^*(A).
\end{equation*}

The quantity $p(B)$ also has combinatorial significance.
Let $G^2$ be a graph with the same vertices as $G$.
Distinct vertices are adjacent in $G^2$ if they are connected by a path of length at most 2 in $G$.
Then the confusability graph of the channel $B$ is $G^2$ and $\alpha(G^2) = p(B)$.

\subsection{Caro-Wei, Motzkin-Straus, and Tur\'{a}n Theorems}
\label{section:turan}
Tur\'{a}n's theorem is 
\begin{equation*}
\alpha(G) \geq \frac{|X|}{1 + \overline{d}(G)},
\end{equation*}
where $\overline{d}(G) = \frac{2|E|}{|X|} = \frac{1}{|X|} \sum_{x \in X} d_G(x)$ is the average degree of $G$.
In this section, we discuss two strengthenings of Tur\'{a}n's theorem: the Caro-Wei theorem and the Motzkin-Straus theorem.
All three of these theorems are often stated in terms of cliques rather than independent sets.
To convert from one form to the other, replace the graph with its complement.

Like $p^*_{\textsc{dsl}}(B,\one)$, the Caro-Wei theorem uses the degree sequence of the graph \cite{alon_turans_2004}.
It states that for a graph $G$,
\begin{equation*}
\alpha(G) \geq \sum_{x \in X} \frac{1}{1 + d_G(x)} = \sum_{x \in X} \frac{1}{(B\one)_x}.
\end{equation*}
The following is an slight generalization of the Caro-Wei theorem.
\begin{lemma}
\label{lemma:caro}
Let $B-I$ be the adjacency matrix of a graph $G$ with vertex set $X$.
For any $t \in R^X$ such that $t \geq \zero$ and $Bt > \zero$, define
\begin{equation*}
\alpha_{\textsc{cw}}(G,t) = \sum_{x \in X} \frac{t_x}{(Bt)_x}.
\end{equation*}
Then $\alpha(G) \geq \alpha_{\textsc{cw}}(G,t)$.
\end{lemma}
\begin{IEEEproof}
Let $h \in \R^X$ be a vector of independent exponentially distributed random variables such that $h_x$ has mean $1/t_x$.
Order the vertices by the value of their entry in $h$.
Let $S(h) = \{x \in X : h_x < \min_{u \in N(x)} h_u\}$.
This set contains the vertices that appear earlier in the ordering than all of their neighbors.
For any $h$, $S(h)$ is an independent set in $G$.
Then 
\begin{equation*}
\E |S(h)| = \sum_{x \in X} \mathrm{Pr}[x \in S(h)] = \sum_{x \in X} \frac{t_x}{(Bt)_x}.
\end{equation*}
Some independent set must be at least as large as $\E |S(h)|$.
\end{IEEEproof}

The Motzkin-Straus theorem is an immediate consequence of Lemma~\ref{lemma:caro}. 
It states that for any $t \in R^X$ such that $t \geq \zero$ and $Bt > \zero$,
\begin{equation*}
\alpha(G) \geq \frac{(\one^T t)^2}{t^T B t}.
\end{equation*}
Call the quantity in this lower bound $\alpha_{\textsc{ms}}(G,t)$.
The function $f(x) = x^{-1}$ is convex, so by Jensen's inequality
\begin{equation*}
\one^T t\sum_{x \in X} \frac{t_x}{\one^T t} (Bt)_x^{-1} \geq \one^T t \left(\sum_{x \in X} \frac{t_x}{\one^T t} (Bt)_x \right)^{-1} = \frac{(\one^T t)^2}{t^T B t}.
\end{equation*}

Specializing to $t = \one$, we obtain Tur\'{a}n's theorem:
\begin{equation*}
\frac{(\one^T \one)^2}{\one^T B \one} = \frac{|X|^2}{|X| + 2|E(G)|} = \frac{|X|}{1 + \frac{2|E(G)|}{|X|}} = \frac{|X|}{1 + \overline{d}(G)}.
\end{equation*}

The expression for Tur\'{a}n's theorem is very similar to the expression for the maximum degree bound $p^*_{\textsc{mdl}}(B, \one)$, except that the maximum degree has been replaced with the average degree.
Thus Tur\'{a}n's theorem is always stronger.
In fact, $p^*_{\textsc{mdl}}(B, t) \leq \alpha_{\textsc{ms}}(G,t)$ for all admissible $t$.
Additionally, the vector $z = \frac{1}{1 + \overline{d}(G)} \one = \frac{|X|}{\one^TB\one}\one$ is feasible in the program for $p^*_{\textsc{dsl}}(B,\one)$, so $p^*_{\textsc{dsl}}(B,\one) \leq \alpha_{\textsc{ms}}(G,\one)$.

The Caro-Wei lower bound $\alpha_{\textsc{cw}}(G,t)$ is always at least as large as the degree sequence lower bound $p^*_{\textsc{dsl}}(B, t)$.
In fact, the Caro-Wei number is always between the local degree lower and upper bounds on $\kappa^*(B)$.
For any $x \in X$,
\begin{equation*}
\min_{y \in N(x)} (Bt)_y \leq (Bt)_x \leq \max_{y \in N(x)} (Bt)_y
\end{equation*}
so
\begin{IEEEeqnarray*}{ClCl}
& \kappa^*_{\textsc{ldl}}(B,t) &=& \sum_{x \in X} \frac{t_x}{\max_{y \in N(x)} (Bt)_y} \\
\leq & \alpha_{\textsc{cw}}(G,t) &=& \sum_{x \in X} \frac{t_x}{(Bt)_x} \\
\leq & \kappa^*_{\textsc{ldu}}(B,t) &=&  \sum_{x \in X} \frac{t_x}{\min_{y \in N(x)} (Bt)_y}.
\end{IEEEeqnarray*}

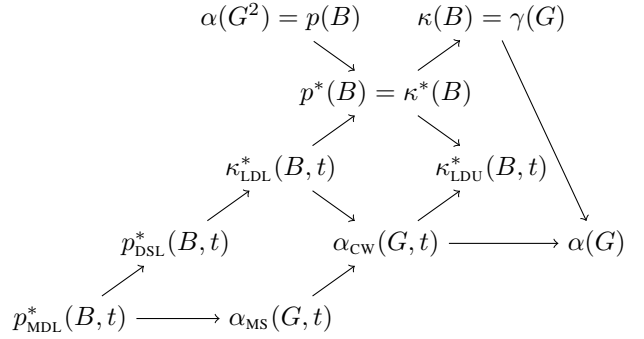
\begin{figure}
\centering
 \begin{tikzpicture}[auto]
 \node (alpha)  at (7.0, 1) {$\alpha(G)$};
 \node (gamma)  at (5.6, 4) {$\kappa(B) = \gamma(G)$};
 \node (star)   at (4.2, 3) {$p^*(B) = \kappa^*(B)$};
 \node (alpha2) at (2.8, 4) {$\alpha(G^2) = p(B)$};
 \node (ldub)   at (5.6, 2) {$\kappa^*_{\textsc{ldu}}(B,t)$};
 \node (ldlb)   at (2.8, 2) {$\kappa^*_{\textsc{ldl}}(B,t)$};
 \node (caro)   at (4.2, 1) {$\alpha_{\textsc{cw}}(G,t)$};
 \node (ms)     at (2.8, 0) {$\alpha_{\textsc{ms}}(G,t)$};
 \node (ds)     at (1.4, 1) {$p^*_{\textsc{dsl}}(B,t)$};
 \node (maxd)   at (  0, 0) {$p^*_{\textsc{mdl}}(B,t)$};

 \draw[->] (maxd) -- (ms);
 \draw[->] (ms) -- (caro);
 \draw[->] (caro) -- (alpha);

 \draw[->] (maxd) -- (ds);
 \draw[->] (ds) -- (ldlb);
 \draw[->] (ldlb) -- (star);
 \draw[->] (star) -- (ldub);

 \draw[->] (alpha2) -- (star);
 \draw[->] (star) -- (gamma);
 \draw[->] (gamma) -- (alpha);

 \draw[->] (ldlb) -- (caro);
 \draw[->] (caro) -- (ldub);
% \draw[->] (0,-0.7) to node {$\leq$} (7,-0.7);
 \end{tikzpicture}
\caption{
Lower bounds on $\alpha(G)$, where $B-I$ is the adjacency matrix of $G$. 
For each arrow, the quantity at the tail is at most as large as the quantity at the head.
If $G$ is regular, then $p^*_{\textsc{mdl}}(B,t) = \kappa^*_{\textsc{ldu}}(B,t)$ so all of the fractional sphere-covering bounds and Tur\'{a}n type bounds to that value.}
\label{figure:inequalities}
\end{figure}

The inequalities among all of these lower bounds on $\alpha(G)$ are summarized in Figure~\ref{figure:inequalities}.
Note that there is no inequality relating the Caro-Wei number to either the domination number or the fractional domination number.
The star graph $K_{1,k}$ demonstrates that the Caro-Wei number can much larger than the domination number: $\alpha_{\textsc{cw}}(K_{1,k}, \one) = \frac{k}{2} + \frac{1}{k+1}$ while $\alpha(K_{1,k}^2) = \gamma(K_{1,k}) = 1$.
For the $n$-vertex path graph $P_n$, the inequality can go in the other direction:
%Note that $\alpha_{\textsc{cw}}(P_{3k}, \one) = \frac{3k-2}{3}+\frac{2}{2} = k + \frac{1}{3}$, but $\alpha(P_{3k}^2) = \gamma(P_{3k}) = k$.
%On the other hand, $\alpha_{\textsc{cw}}(P_{3k+1}, \one) = \frac{3k-1}{3}+\frac{2}{2} = k + \frac{2}{3}$, but $\alpha(P_{3k+1}^2) = \gamma(P_{3k+1}) = k+1$.
%(Also $\alpha_{\textsc{cw}}(P_{3k+2}, \one) = \alpha(P_{3k+2}^2) = \gamma(P_{3k+2}) = k+1$.)

\begin{tabular}{|l||l|l|}
\hline
$G$ & $\alpha_{\textsc{cw}}(G, \one)$ & $\alpha(G^2) = \gamma(G)$\\
\hline
$P_{3k}$   & $k + \frac{1}{3}$ & $k$ \\
$P_{3k+1}$ & $k + \frac{2}{3}$ & $k + 1$ \\
$P_{3k+2}$ & $k + 1$           & $k + 1$ \\
\hline
\end{tabular}

For the strong graph product of $n$ copies of $P_4$, the Caro-Wei number is $\left(\frac{5}{3}\right)^n$ and the domination number is $2^n$.
Thus the gap between the two bounds can be arbitrarily large in either direction.
The Caro-Wei number is always larger than any of the approximations to the fractional domination number presented in this paper.
If the exact value fractional domination number cannot be efficiently computed, the Caro-Wei number is likely to be the best available lower bound.

\subsection{Bounds using only the number of edges}
Tur\'{a}n's theorem uses the number of edges in $G$, or equivalently the number of edges in the channel graph of $B$, to give a reasonably good lower bound on $\alpha(G) = p(A)$.
However, the best lower bound on $p(A)$ that uses only the number of edges in the channel graph of $A$ is very weak.
\begin{lemma}
\label{lemma:edge-only-lb}
Let $A \in \mtrx{\two}{X}{Y}$ be a channel and let $E$ be the edge set of the channel graph. Then
\begin{equation}
|X| + |Y| - |E| \leq p(A). \label{eq:eolb}
\end{equation}
For any $X$, $Y$, and $R \subseteq Y$ such that $|R| \leq |X|$, there is a channel $A$ such that $|E| =|X| + |Y| - |R|$ and $R$ is an output covering in $A$. 
Thus \eqref{eq:eolb} cannot be improved.
\end{lemma}
\begin{IEEEproof}
For each $y \in Y$, we select $|N(y)| - 1$ inputs to forbid from the code.
We forbid at most $|E| - |Y|$ total inputs, so our code contains at least $|X| + |Y| - |E|$ inputs.

We construct the tightness example $A$ as follows.
Choose the neighborhoods of the outputs in $R$ so that each is nonempty, they are disjoint, and $\bigcup_{y \in R} N(y) = X$.
Meeting the first two conditions is possible because $|R| \leq |X|$.
Because the union of neighborhoods cover all of $X$, $R$ is a covering and $p(A) \leq |R|$.
We have included $|X|$ edges so far.
For each $y \in Y \setminus R$, let $|N(y)| = 1$ and choose the neighbor arbitrarily.
Thus $|E| = |X| + |Y| - |R|$.
\end{IEEEproof}
Only a few edges are needed to create a small number of output vertices with large degree.
Compare this to Lemma~\ref{lemma:edge-only-ub}.
For the tightness example $A$, note that the channel graph of $B = A \circ A$ has $\sum_{y \in R} N(y)^2$ edges.
This is at least $\frac{|X|^2}{|R|}$, which is usually much larger than $|X|+|Y|-|R|$ and is also exactly the number of edges forced by Tur\'{a}n's theorem.

\section{Conclusion}
We have discussed a wide variety of upper and lower bounds on the size of codes for combinatorial channels.
We can summarize the most important points as follows.
When the channel is input-regular, the minimum degree, degree sequence, and local degree upper bounds here are equivalent, but not necessarily equal to the fractional covering number.
When the channel is also output-regular, all of the upper bounds discussed become equivalent.
Knowledge of the average input degree alone gives a very weak bound.
If the input degrees concentrate around the average, the degree sequence bound will be fairly strong.
The local degree bound is always at least as good as the degree sequence bound but uses more information about the structure of the channel.
The local degree bound can be iterated to obtain stronger bounds.
The best sphere packing bound for a given channel can be much weaker than the best sphere packing bound for some other channel that admits the same codes.
Consequently, finding a family of channels equivalent to the channel of interest can be very powerful.
If the confusion graph is regular, all lower bound discussed in this paper are equivalent.
In contrast to the situation for upper bounds, knowledge of the average degree alone is sufficient to obtain a good lower bound.
The best of the lower bounds discussed are the Caro-Wei bound and the fractional dominating set number.
These bounds are incomparable in general, but the Caro-Wei bound is better than the local degree lower bound for fractional dominating set.

\bibliographystyle{IEEEtran} 
\bibliography{IEEEabrv,hand-edited}

\appendices

\section{Proofs}

\newtheorem*{T1}{Theorem \ref{thm:1d-feasible-pt}}
\newtheorem*{T2}{Theorem \ref{thm:1dub}}
\newtheorem*{T3}{Theorem \ref{thm:1g-feasible-pt}}

\newtheorem*{L1}{Lemma \ref{lemma:threshold}}
\newtheorem*{L4}{Lemma \ref{lemma:feasible-pt-three}}

\begin{T1}
\thmonetext
\end{T1}
\begin{IEEEproof}
By Lemma~\ref{lemma:local-iteration}, $\varphi \circ \varphi(\one)$ is feasible for $\kappa^*(A_n)$.
From the definition of $\varphi$,
\begin{equation*}
\frac{z_y}{\varphi(z)_y} = \min_{x \in N(y)} (A_nz)_x
\end{equation*}

Each $x \in [2]^n$ has $r_x$ total substrings, so $(A_nz'')_x = r_x$,  
\begin{equation*}
\frac{1}{\varphi(\one)_y} = \min_{x \in N(y)} (A_n \one)_x = \min_{x \in N(y)} r_x = r_y,
\end{equation*}
and $\varphi(\one)_y = 1/r_y$.

Of the substrings of $x$, $u_x-b_x$ have $r_x-2$ runs, $b_x$ have $r_x-1$ runs, and $r_x-u_x$ have $r_x$ runs, so
\begin{IEEEeqnarray*}{rCl}
&&(A_n\varphi(\one))_x \\
&=& \sum_{y \in N(x)} \frac{1}{r_y}\\
&=& \frac{u_x - b_x}{r_x-2} +\frac{b_x}{r_x-1} +\frac{r_x-u_x}{r_x}\\
&=& 1 + u_x\left(\frac{1}{r_x-2} - \frac{1}{r_x}\right) + b_x\left(\frac{1}{r_x-1} - \frac{1}{r_x-2}\right)\\
&=& 1 + \frac{2u_x(r_x-1) - b_xr_x}{r_x(r_x-1)(r_x-2)}\\
&=& 1 + \frac{(2u_x - b_x)(r_x-2) +2(u_x - b_x)}{r_x(r_x-1)(r_x-2)}\\
&\geq& 1 + \frac{2u_x - b_x}{r_x(r_x-1)}.
\end{IEEEeqnarray*}
The inequality follows from $u_x - b_x \geq 0$.

Let $y \in [2]^{n-1}$ be a string and let $x \in [2]^n$ be a superstring of $y$.
It is possible to create a superstring by extending an existing run, adding a new run at an end of the string, or by splitting an existing run into three new runs, so $r_x \leq r_y + 2$
The only way to destroy a unit run in $y$ is to extend it into a run of length two, so $u_x \geq u_y-1$.
Similarly, $u_x - b_x \geq u_y - b_y -1$, so $2u_x - b_x \geq 2u_y - b_y -2$.
Applying these inequalities to $(A_n\varphi(\one))_x$, we conclude that
\begin{IEEEeqnarray*}{rCl}
\frac{\varphi(\one)_y}{(\varphi \circ \varphi(\one))_y} &=& \min_{x \in N(y)} (A_n\varphi(\one))_x \\
&\geq& 1 + \frac{\max (2u - b - 2,0)}{(r_y+2)(r_y+1)},\\
(\varphi \circ \varphi(\one))_y &\leq& \frac{1}{r_y} \left( 1 + \frac{\max (2u - b - 2,0)}{(r_y+2)(r_y+1)} \right)^{-1}.
\end{IEEEeqnarray*}
\end{IEEEproof}
\begin{lemma}
\label{lemma:string-count}
The number of strings in $[2]^n$ with $r$ runs is $2\binom{n-1}{n-r}$.

The number of strings in $[2]^n$ with $r$ runs and $u$ unit runs is $2 \binom{n-r-1}{n-2r+u} \binom{r}{r-u}$.

For $r \geq 2$, the number of strings in $[2]^n$ with $r$ runs, $u$ unit runs and $b$ external unit runs is 
$2 \binom{n-r-1}{n-2r+u} \binom{r-2}{u-b} \binom{2}{b}$.
\end{lemma}
\begin{IEEEproof}
For $k \geq 1$, there are $\binom{n+k-1}{n}$ ways to partition $n$ identical items into $k$ distinguished groups.
Thus there are $\binom{n-lk+k-1}{n-lk} = \binom{n-(l-1)k-1}{n-lk}$ ways to partition $n$ items into $k$ groups such that each group contains at least $l$ items.

A binary string is uniquely specified by its first symbol and it run length sequence.
We have $n$ symbols to distribute among $r$ runs such that each run contains at least one symbol, so there are $\binom{n-1}{n-r}$ arrangements.
This proves the first claim.
We can also specify the run sequence of a string by giving the locations of the unit runs and the lengths of the longer runs.
The $r-u$ runs of length at least two can appear in $r$ positions so there are $\binom{r}{r-u}$ arrangements, 
We have $n-u$ symbols to distribute among $r-u$ runs such that each run contains at least 2 symbols, so there are $\binom{n-u - (r-u)-1}{n-u-2(r-u)} = \binom{n-r-1}{n-2r+u}$ arrangements, which proves the second claim.
As long as $r \geq 2$, the internal unit runs two can appear in $r-2$ positions and the external unit runs can appear in 2 positions, so there are $\binom{r-2}{u-b}\binom{2}{b}$ possible arrangements, which proves the third claim.
\end{IEEEproof}
Note that Lemma~\ref{lemma:string-count} uses the polynomial definition of binomial coefficients, which can be nonzero even when the top entry is negative.
For example, the number of strings of length $n$ with $n$ runs, $n$ unit runs, and 2 external unit runs is $2 \binom{-1}{0} \binom{n-2}{n-2} \binom{2}{2} = 2$.

For compactness, let
\begin{equation*}
\E_r [f(r)] = \frac{1}{2^{n-1}} \sum_{r \geq 1} \binom{n-1}{n-r} f(r).
\end{equation*}
and let $\displaystyle \E_{u,b} [f(r,u,b)]$ equal
\begin{equation*}
\frac{1}{\binom{n-1}{n-r}} \sum_{u=0}^r \sum_{b=0}^2 \binom{n-r-1}{n-2r+u}\binom{r-2}{u-b}\binom{2}{b}f(r,u,b)
\end{equation*}
for $r> 1$ and let $\E_{u,b} [f(1,u,b)] = f(1,0,0)$.

If $z_x = f(r_x,u_x,b_x)$, then 
\begin{IEEEeqnarray}{rCl}
\one^T z
&=&\sum_{x \in [2]^n} f(r_x,u_x,b_x) \nonumber\\
&=& 2 f(1,0,0) + \nonumber\\
& & 2\sum_{r=2}^n \sum_{u=0}^r \sum_{b=0}^2 \binom{n-r-1}{n-2r+u}\binom{r-2}{u-b}\binom{2}{b}f(r,u,b) \nonumber\\
&=& 2^n \E_r\left[\E_{u,b}\left[f(r,u,b)\right]\right] \label{eq:unit-run-sum}
\end{IEEEeqnarray}

Analysis of the feasible point constructed in Theorem~\ref{thm:1d-feasible-pt} relies on the following identities.
For $k \geq 0$,
\begin{IEEEeqnarray}{rCl}
\E_r \left[ \frac{1}{\binom{r+k-1}{r-1}} \right] 
&=& \frac{\sum_{r \geq 1} \binom{n+k-1}{n-r}}{2^{n-1}\binom{n+k-1}{n-1}} \nonumber \\
&\leq& \frac{2^k}{\binom{n+k-1}{n-1}} \label{eq:r}\\
\E_{u,b} \left[ \binom{u}{u-k} \right] &=& \frac{\binom{r}{r-k} \binom{r-1}{r-k-1}}{\binom{n-1}{n-k-1}} \label{eq:u}\\
\E_{u,b} \left[ b \right] &=& \frac{2 (r-1)}{n-1}.  \label{eq:b}
\end{IEEEeqnarray}
Each of these can be easily derived from the binomial theorem and Vandermonde's identity.

\begin{T2}
For $n \geq 2$,
\begin{equation*}
\kappa^*(A_n) \leq \frac{2^n}{n+1}\left(1 + \frac{26}{n(n-1)}\right).
\end{equation*}
\end{T2}
\begin{IEEEproof}
For $n \leq 13$, this follows from the bound of Kulkarni and Kiyavash \cite{kulkarni_nonasymptotic_2013}.
This proof covers $n \geq 10$.

From Theorem~\ref{thm:1d-feasible-pt} and \eqref{eq:unit-run-sum}, we have $\kappa^*(A_{n+1}) \leq 2^n \E_r [ \E_{u,b} [ f(r,u,b)]]$ where
\begin{IEEEeqnarray*}{rCl}
f(r,u,b) &=& \frac{1}{r} \left(1 + \frac{\max(2u - b - 2,0)}{(r+2)(r+1)}\right)^{-1}.
\end{IEEEeqnarray*}

For $x > 0$, $(1+x)^{-1} \leq 1-x+x^2$, so $f(r,u,b)$ is at most
\begin{IEEEeqnarray*}{Cl}
    & \frac{1}{r} \left(1 - \frac{\max(2u - b - 2,0)}{(r+2)(r+1)} + \frac{(\max(2u - b - 2,0))^2}{(r+2)^2(r+1)^2}\right)\\
\leq& \frac{1}{r} - \frac{2u - b - 2}{(r+2)(r+1)r} + \frac{2u(2u-2)}{(r+2)^2(r+1)^2r}
\end{IEEEeqnarray*}
We will bound this term by term using \eqref{eq:r}, \eqref{eq:u}, and \eqref{eq:b}.
First
\begin{IEEEeqnarray*}{Cl}
&\E_r \left[ \E_{u,b} \left[ \frac{1}{r} - \frac{2u - b - 2}{(r+2)(r+1)r} \right] \right] \\
=& \E_r \left[ \frac{1}{r} - \frac{1}{(r+2)(r+1)r} \left( 2 \frac{r(r-1)}{n-1} - \frac{2(r-1)}{n-1} - 2\right) \right] \\
=& \E_r \left[ \frac{1}{r} - \frac{2}{n-1} \cdot \frac{(r-1)^2 - n + 1}{(r+2)(r+1)r} \right] \\
%=& \frac{2}{n-1} \E \left[ \frac{n-1}{2r} -  \frac{r^2 + 3r + 2 - (5r + 10) + 10 - n}{(r+2)(r+1)r} \right] \\
=& \frac{2}{n-1} \E_r \left[ \frac{n-1}{2r} -  \frac{1}{r} + \frac{5}{(r+1)r} + \frac{n - 10}{(r+2)(r+1)r} \right] \\
=& \frac{2}{n-1} \E_r \left[ \frac{n-3}{2r} + \frac{5}{(r+1)r} + \frac{n - 10}{(r+2)(r+1)r} \right] \\
\leq& \frac{2}{n-1} \left( \frac{n-3}{n} + \frac{20}{(n+1)n} + \frac{8n - 80}{(n+2)(n+1)n} \right) \\
=& \frac{2}{n-1} \left( \frac{n^2-n-6}{(n+2)n} + \frac{28n - 40}{(n+2)(n+1)n} \right) \\
=& \frac{2}{n+2} \left( \frac{n^2-n-6}{n(n-1)} + \frac{28n - 40}{(n+1)n(n-1)} \right) \\
=& \frac{2}{n+2} \left( 1 + \frac{22n - 46}{(n+1)n(n-1)} \right) \\
<& \frac{2}{n+2} \left( 1 + \frac{22}{(n+1)n} \right).
\end{IEEEeqnarray*}
Second,
\begin{IEEEeqnarray*}{Cl}
    & \E_r \left[\E_{u,b} \left[\frac{4u(u-1)}{(r+2)^2(r+1)^2r} \right] \right] \\
   =& \E_r \left[\frac{4}{(r+2)^2(r+1)^2r} \cdot \frac{r(r-1)^2(r-2)}{(n-1)(n-2)} \right] \\
  <& \E_r \left[\frac{4}{(r+2)(r+1)} \cdot \frac{(r-1)(r-2)}{(n-1)(n-2)} \cdot \frac{1}{r}  \right] \\
\leq& \E_r \left[\frac{4}{(r+2)(r+1)} \cdot \frac{(r+2)(r+1)}{(n+2)(n+1)} \cdot \frac{1}{r}  \right] \\
   =& \E_r \left[\frac{4}{(n+2)(n+1)} \cdot \frac{1}{r} \right]\\
\leq& \frac{8}{(n+2)(n+1)n}.
\end{IEEEeqnarray*}
Combining these two terms, we get
\begin{equation*}
\kappa^*(A_{n+1}) \leq 2^n \frac{2}{n+2} \left( 1 + \frac{22}{(n+1)n} + \frac{4}{(n+1)n}\right).
\end{equation*}

\end{IEEEproof}

\begin{L4}
Let $z_y = f'(r_y,u_y,b_y)$. Then 
\begin{equation*}
\one^T z \geq \frac{2^n - 2}{n+1}\left(1+\frac{1}{n-1} - \frac{3}{(n-1)(n-2)}\right)
\end{equation*}
\end{L4}
\begin{IEEEproof}
\begin{IEEEeqnarray*}{rCl}
f'(r,u,b) &\geq& \frac{1}{r} \left(1 - \frac{u-b}{r^2}\right)\\
  &\geq& \frac{1}{r} \left(1 - \frac{u-b}{(r-1)(r-2)}\right)\\
\end{IEEEeqnarray*}

\begin{IEEEeqnarray*}{Cl}
 &2^n\E_r \left[ \E_{u,b} \left[\frac{1}{r} \left(1 - \frac{u-b}{(r-1)(r-2)}\right)\right]\right]\\
=&2^n\E_r \left[\frac{1}{r} \left(1 - \frac{1}{(r-1)(r-2)}\left( \frac{r(r-1)}{n-1} - \frac{2(r-1)}{n-1} \right)\right)\right]\\
=&2^n\E_r \left[\frac{1}{r} \left(1 - \frac{1}{n-1}\right)\right]\\
=&2^n\frac{2^n - 1}{2^{n-1}n} \left(1 - \frac{1}{n-1}\right)\\
=&\frac{2^{n+1} - 2}{n+2} \left(\frac{(n+2)(n-2)}{n(n-1)}\right)\\
=&\frac{2^{n+1} - 2}{n+2} \left(1 + \frac{1}{n} - \frac{3}{n(n-1)}\right).
\end{IEEEeqnarray*}
\end{IEEEproof}

\begin{T3}
Let $A_n$ be the $n$-bit 1-grain-error channel.
The vector 
\begin{equation*}
z_y = \frac{1}{r_y} \left(1 + \frac{2u_y - 2b_y^R - b_y^L - 2}{(r_y+2)(r_y+1)}\right)^{-1}
\end{equation*}
is feasible for $\kappa^*(A_n)$.
\end{T3}
\begin{IEEEproof}
By Lemma~\ref{lemma:local-iteration}, $\varphi \circ \varphi(\one)$ is feasible for $\kappa^*(A)$.
From the definition of $\varphi$,
\begin{equation*}
\frac{z_y}{\varphi(z)_y} = \min_{x \in N(y)} (A_nz)_x
\end{equation*}

Each $x \in [2]^n$ has $r_x$ total neighbors, so $(A_nz'')_x = r_x$,  
\begin{equation*}
\frac{1}{\varphi(\one)_y} = \min_{x \in N(y)} (A \one)_x = \min_{x \in N(y)} r_x = r_y,
\end{equation*}
and $\varphi(\one)_y = 1/r_y$.

Of the neighbors of $x$, $u_x-b_x^L-b_x^R$ have $r_x-2$ runs, $b_x^L$ have $r_x-1$ runs, and $r_x-u_x+b_x^R$ have $r_x$ runs, so $(A_n\varphi(\one))_x$ equals
\begin{IEEEeqnarray*}{Cl}
& \sum_{y \in N(x)} \frac{1}{r_y}\\
=& \frac{u_x - b_x^L - b_x^R}{r_x-2} +\frac{b_x^L}{r_x-1} +\frac{r_x-u_x+b_x^R}{r_x}\\
=& 1 + (u_x - b_x^R)\left(\frac{1}{r_x-2} - \frac{1}{r_x}\right) + b_x^L\left(\frac{1}{r_x-1} - \frac{1}{r_x-2}\right)\\
=& 1 + \frac{2(u_x - b_x^R)(r_x-1) - b_x^Lr_x}{r_x(r_x-1)(r_x-2)}\\
=& 1 + \frac{(2u_x - 2b_x^R - b_x^L)(r_x-2) + 2(u_x - b_x^R - b_x^L)}{r_x(r_x-1)(r_x-2)}\\
\geq& 1 + \frac{2u_x - 2b_x^R - b_x^L}{r_x(r_x-1)}.
\end{IEEEeqnarray*}

Let $x \in [2]^n$ be an input and let $y \in N(x)$.
A grain error can leave the number of runs unchanged, destroy a unit run at the start of $x$, or destroy a unit run in the middle of $x$, merging the adjacent runs.
Thus $r_y \geq r_x - 2$
The only way to produce a unit run in $y$ is shorten a run of length two in $x$, so $u_x \geq u_y-1$.
Similarly, $2u_x - 2b_x^R - b_x^L \geq 2u_y - 2b_y^R - b_y^L -2$.
Applying these inequalities to $(A\varphi(\one))_x$, we conclude that
\begin{IEEEeqnarray*}{rCl}
\frac{\varphi(\one)_y}{(\varphi \circ \varphi(\one))_y} &=& \min_{x \in N(y)} (A\varphi(\one))_x \geq 1 + \frac{2u_y - 2b_y^R - b_y^L - 2}{(r_y+2)(r_y+1)},\\
(\varphi \circ \varphi(\one))_y &\leq& \frac{1}{r_y} \left( 1 + \frac{2u_y - 2b_y^R - b_y - 2}{(r_y+2)(r_y+1)} \right)^{-1}.
\end{IEEEeqnarray*}

\end{IEEEproof}

\begin{L1}
$\kappa^*(A_{q,n,a,b}) \leq \kappa^*(A_{q,n,a+2,b-1})$ when $a + qb \leq n-1$.
\end{L1}
\begin{IEEEproof}
We can rewrite the initial inequality as
\begin{IEEEeqnarray}{rCl}
\kappa^*(A_{q,n,a+2,b-1}) &\geq& \kappa^*(A_{q,n,a,b}) \nonumber\\
\frac{q^{n-a-2}}{\sum_{i=0}^{b-1} \binom{n-a-2}{i}(q-1)^i} &\geq& \frac{q^{n-a}}{\sum_{i=0}^b \binom{n-a}{i}(q-1)^i} \nonumber\\
\sum_{i=0}^{b} \binom{n-a}{i} (q-1)^i &\geq& q^2 \sum_{i=0}^{b-1} \binom{n-a-2}{i} (q-1)^i \label{eq:difference}
\end{IEEEeqnarray}
To simplify \eqref{eq:difference}, we us the following identity:
\begin{IEEEeqnarray*}{Cl}
&\sum_{i=0}^{b} \binom{n-c+2}{i} (q-1)^i \\
=&\sum_{i=0}^{b} \left(\binom{n-c}{i-2} + 2 \binom{n-c}{i-1} + \binom{n-c}{i}\right)(q-1)^i \\
=&\sum_{i=0}^{b-2} \binom{n-c}{i} (q-1)^{i+2} + 2\sum_{i=0}^{b-1} \binom{n-c}{i} (q-1)^{i+1} + \\
&\quad \sum_{i=0}^{b} \binom{n-c}{i} (q-1)^i \\
=&\binom{n-c}{b}(q-1)^{b} - \binom{n-c}{b-1}(q-1)^{b+1} + \\
&\quad ((q-1)^2 + 2(q-1) +1)\sum_{i=0}^{b-1} \binom{n-c}{i} (q-1)^i\\
=& \binom{n-c}{b-1}(q-1)^b \left( \frac{n-c-b+1}{b} - q + 1\right) + \\
&\quad q^2\sum_{i=0}^{b-1} \binom{n-c}{i} (q-1)^i.
\end{IEEEeqnarray*}
By setting $c=a+2$, we can use this to rewrite the left side of \eqref{eq:difference}.
Eliminating the common term from both sides of the inequality gives
\begin{IEEEeqnarray*}{rCl}
\binom{n-a-2}{b-1}(q-1)^b \left( \frac{n-a-b-1}{b} - q + 1\right) &\geq & 0\\
\frac{n-a-b-1}{b} - q + 1 &\geq & 0\\
n-a-1- qb &\geq & 0
\end{IEEEeqnarray*}
which proves the claim.
\end{IEEEproof}

\end{document}